\documentclass[pdflatex,sn-nature]{sn-jnl}

\usepackage{graphicx}%
\usepackage{multirow}%
\usepackage{amsmath,amssymb,amsfonts}%
\usepackage{amsthm}%
\usepackage{mathrsfs}%
\usepackage[title]{appendix}%
\usepackage{xcolor}%
\usepackage{textcomp}%
\usepackage{manyfoot}%
\usepackage{booktabs}%
\usepackage{algorithm}%
\usepackage{algorithmicx}%
\usepackage{algpseudocode}%
\usepackage{listings}%
\usepackage{pdflscape}  
\usepackage{CJKutf8}
\usepackage{csvsimple}
\usepackage{tabularx}
\usepackage{float}
\usepackage{pdfpages}

\theoremstyle{thmstyleone}%
\theoremstyle{thmstyletwo}%

\theoremstyle{thmstylethree}%

\begin{document}

\title[Article Title]{{Stable Emotional Co-occurrence Patterns Revealed by Network Analysis of Social Media}}

\author[1]{\fnm{Qianyun} \sur{Wu}}\email{maggie.w.aa@m.titech.ac.jp}

\author*[2]{\fnm{Orr} \sur{Levy}}\email{orr.levy@gmail.com}

\author*[3]{\fnm{Yoed} \sur{N. Kenett}}\email{yoedk@technion.ac.il}

\author[4]{\fnm{Yukie} \sur{Sano}}\email{sano@sk.tsukuba.ac.jp}

\author[1]{\fnm{Hideki} \sur{Takayasu}}\email{takayasu.h.aa@m.titech.ac.jp}

\author*[1,5]{\fnm{Shlomo} \sur{Havlin}}\email{havlins@gmail.com}

\author*[1]{\fnm{Misako} \sur{Takayasu}}\email{takayasu@comp.isct.ac.jp}

\affil[1]{\orgdiv{School of Computing}, \orgname{Institute of Science Tokyo}, \orgaddress{\city{Yokohama}, \postcode{226-8501}, \country{Japan}}}

\affil[2]{\orgdiv{Faculty of Engineering}, \orgname{Bar-Ilan University}, \orgaddress{\city{Ramat-Gan}, \postcode{5290002}, \country{Israel}}}

\affil[3]{\orgdiv{Faculty of Data and Decision Sciences}, \orgname{Technion – Israel Institute of Technology}, \orgaddress{\city{Haifa}, \postcode{3200003}, \country{Israel}}}

\affil[4]{\orgdiv{Institute of Systems and Information Engineering}, \orgname{University of Tsukuba}, \orgaddress{\city{Ibaraki}, \postcode{305-8577}, \country{Japan}}}

\affil[5]{\orgdiv{Department of Physics}, \orgname{Bar-Ilan University}, \orgaddress{\city{Ramat-Gan}, \postcode{5290002},\country{Israel}}}



\abstract {Examining emotion interactions as an emotion network in social media offers key insights into human psychology, yet few studies have explored how fluctuations in such emotion network evolve during crises and normal times. This study proposes a novel computational approach grounded in network theory, leveraging large-scale Japanese social media data spanning varied crisis events (earthquakes and COVID-19 vaccination) and non-crisis periods over the past decade. Our analysis identifies and evaluates links between emotions through the co-occurrence of emotion-related concepts (words), revealing a stable structure of emotion network across situations and over time at the population level. We find that some emotion links (represented as link strength) such as emotion links associated with Tension are significantly strengthened during earthquake and pre-vaccination periods. However, the rank of emotion links remains highly intact. These findings challenge the assumption that emotion co-occurrence is context-based and offer a deeper understanding of emotions' intrinsic structure. Moreover, our network-based framework offers a systematic, scalable method for analyzing emotion co-occurrence dynamics, opening new avenues for psychological research using large-scale textual data.}


\maketitle

\section{. Introduction}\label{sec1}

In recent years, social media has become a vital platform for disseminating information and providing social support during crises such as natural disasters and pandemics \cite{bib1,bib2,bib3}. Crises dramatically impact one’s well-being \cite{bib41}, yet elucidating how one’s emotions may drastically alter in crisis modes is quite challenging in standard lab settings \cite{bib40}, because real-life emotional intensity, unpredictability of crises, and ethical constraints cannot be accurately replicated in controlled environments. Understanding the underlying complexity of users' emotions and how the emotional interaction network structure is affected by crises offers critical insights for policy makers and mental health professionals, enabling the development of timely and effective intervention strategies to support individuals in coping with the psychological challenges that arise in such situations. Our study might also help to emotional readiness of dealing with complex situations, to enable a robust and resilient society.

Emotions are complicated phenomena that encompass various concepts, such as subjective feelings, cognitive concepts, and expressions \cite{bib46}. While traditional research proposed a few basic emotions (i.e. anger, fear, joy \cite{bib42}) and suggested that individuals experience emotions one at a time \cite{bib42,bib43,bib44}, a growing body of evidence indicates that multiple emotions can co-exist within individuals \cite{bib12,bib10,bib3,bib11,bib45}. Moreover, people may experience complex emotional states during crises \cite{bib6,bib22}. For example, during the COVID-19 pandemic, individuals might feel depressed by recurring waves of outbreaks while worrying about the safety of newly developed vaccines. Likewise, in the aftermath of an earthquake, a person may experience fear for their safety alongside anger directed at perceived mismanagement by authorities. However, despite growing interest in the complexity of emotions and efforts to understand how emotions co-occur across different situations, there is a lack of studies examining the dynamic and persistence of emotion co-occurrence within individuals during both crises and normal times—particularly at the scale of large populations.


In the current study, we investigate the evolving emotion network based on the co-occurrence of emotions within individual users across a large population, using a computational data-driven approach, utilizing large-scale social media datasets spanning various crisis events and normal times. Specifically, we address the following two research questions: (1) Are there any fundamental emotional co-occurrence patterns of the online population or users that remain stable (i.e., persistent) over a long period? (2) How does this emotional structure change during crises compared to non-crisis events? Our findings indicate that while the strength of emotional links are persistent within crisis and normal periods, they are different in crisis periods and normal times. However, their relative ranking, as assessed using Spearman's correlation-remains highly stable across both temporal scales and situational contexts.

Previous studies examining emotion co-occurrence primarily rely on surveys with relatively small sample sizes (e.g. 21,678 participants \cite{bib10}, or 500 participants \cite{bib47}). These studies typically create experimental scenarios designed to elicit mixed emotional states, followed by self-reported surveys to capture participants’ responses \cite{bib8,bib9,bib10,bib11,bib12}, which may lack enough statistics to draw robust conclusions \cite{bib68}. Furthermore, it has been pointed out that emotion research, for pragmatic reasons, has tended to overlook the fundamentally dynamic nature of emotions \cite{bib27}. Likewise, most emotion co-occurrence research is static. For instance, Moeller et al. \cite{bib10} surveyed 21,678 high school students and built a co-occurrence network from their self-reported emotions, showing that positive and negative feelings—like stress and happiness—often co-occur within individuals. However, there are only very few studies based on small-scale datasets that evaluate how the emotion co-occurrence structures dynamically change over time \cite{bib21,bib60}. Studies with small sample sizes generally suggest that relationships between emotions vary depending on the situation.

Moreover, there is a lack of research on emotional response to crisis \cite{bib26,bib6,bib22}, especially on the network structure of co-occurring emotions during crisis. Previous research on emotional responses to crises has primarily focused on isolated emotional states at the collective level by constructing time series and analyzing the temporal trends of each emotion separately \cite{bib4,bib5,bib7,bib24}. For example, Wu et al. \cite{bib5} found that during the COVID-19 outbreak in China, anger, depression, vigor, and tension often showed "bursty peaks," with vigor more likely triggered by external events, and tension more likely triggered by information spread among users. Sano et al. \cite{bib4} found that sharp spikes in emotions (i.e. tension, depression and confusion) could be associated with natural disasters such as earthquakes. There are only very few studies that examine the relationships between emotions \cite{bib66} which found only for this specific study that fear is positively correlated to reproach and negatively correlated to distress.


Overall, these previous studies highlight a significant gap: the lack of large-scale, systematic investigations into emotion co-occurrence network structures at the population level, especially how these structures might vary during times of crisis compared to non-crisis times.

To bridge these gaps and address the research questions outlined above, we conduct our analysis on a large-scale Japanese social media data spanning 13 years and 47.6 million posts. The dataset includes posts related to major crisis events such as earthquakes and COVID-19 (vaccination), and we compare it to content reflecting everyday non-crises life. We characterize the emotional aspect of our data, by adopting an emotion classification framework based on the Profile of Mood States (POMS) \cite{bib13}, which categorizes emotions into six dimensions: Tension-Anxiety, Depression-Dejection, Anger-Hostility, Vigor-Activity, Fatigue-Inertia, and Confusion-Bewilderment (in the following, we refer to them as Tension, Depression, Anger, Vigor, Fatigue and Confusion). Each dimension consists of multiple emotion concepts—for example, "grouchy" and "annoyed" characterize Anger, while "nervous" and "panicky" represent Tension. This psychological rating system, which primarily focuses on negative emotions, has been widely used to assess mental states before and after significant events such as competitions or surgeries \cite{bib15,bib16,bib48,bib49,bib50,bib51}. Given its sensitivity to emotional changes in response to external stressors, it is well-suited for our study, which aims to capture nuanced emotional profiles during crises. The POMS includes several emotions - Tension, Depression, Anger, and Vigor - that are similar to other widely used emotion models such as Ekman's six basic emotions~\cite{bib42} (anger, disgust, fear, happiness, sadness, surprise) and Plutchik's wheel \cite{bib36} (joy, sadness, trust, disgust, fear, anger, surprise, and anticipation). The other two emotions—Fatigue and Confusion—though less commonly used in other emotion models, are considered as important and complex emotions that impact individuals' conscious experiences and performance \cite{bib58,bib59}.


Here, we develop a novel network framework for systematically measuring and comparing emotion co-occurrence in large social media. Network-based methods have recently gained significant attention in psychology and cognitive science \cite{bib52,bib53}, as an effective tool for examining the complex interrelationships among emotional \cite{bib10,bib9,bib27,bib28,bib29,bib47,bib63,bib64}, semantic \cite{bib30,bib32,bib54,bib55}, and studying personality \cite{bib33,bib34,bib56} concepts. We hypothesize here that the mixture of emotions expressed by an individual at a given post, can be represented through links between emotion concepts (words) used within this post (see demonstration in \textbf{Fig. 1}). To capture patterns of emotion co-occurrence at the group level, we begin with constructing an emotional concept network by aggregating pairs of co-occurring emotion concepts across multiple social media posts. In this network, each node represents a POMS concept (word) and each link denotes a significant co-occurrence between two concepts. Significance is measured as the extent to which the observed frequency of a link exceeds its expected frequency under random coexistence (see the definition of significance in \textit{Methods} section). Next, we aggregate these significant emotion concept links to build an emotion network that captures the strength of connections within and among the six POMS emotion dimensions, under the assumption that two emotions co-occur more strongly when they are connected by a larger number of significant concept links relative to random. This framework enables systematic analysis and tracking of changes in emotional structure across different contexts and time periods.


\begin{figure}[H]
\centering
\includegraphics[width=0.9\textwidth]{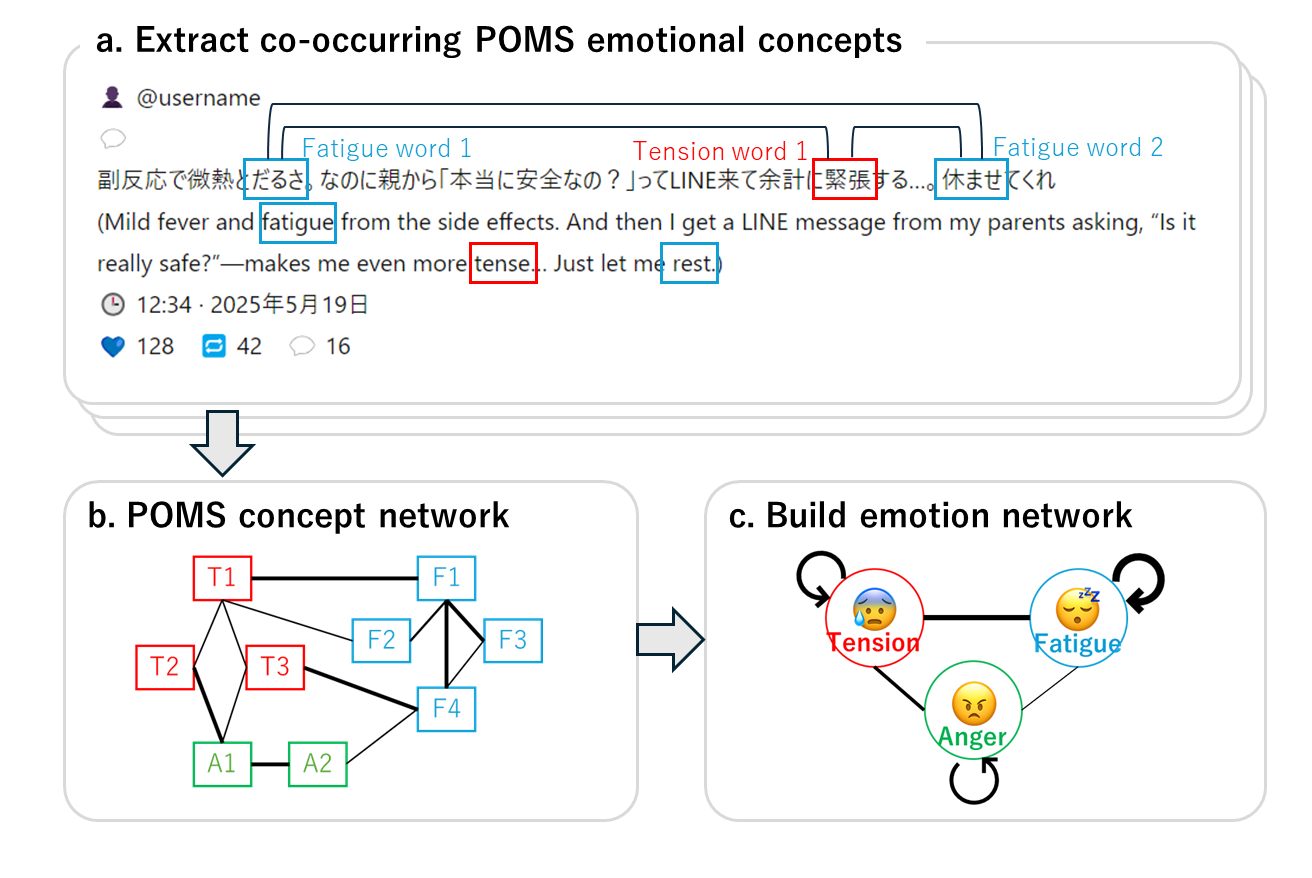}
\caption{Conceptual illustration of steps to build an emotional co-occurrence network based on co-occurring POMS concepts within a single post. (a) Step 1: Extract pairs of POMS concepts that co-exist within the same post. (b) Step 2: Construct a POMS concept network. After aggregating the co-occurring pairs of POMS concepts (words) over multiple posts, we get a network where each node represents an emotion concept, and each link strength represents the significance of co-occurrence between emotion concepts (see \textit{Methods}). (c) Step 3: Build an emotion network by aggregating the number of significant concept links between and within emotion nodes.}\label{fig1}
\end{figure}

\section{Results}\label{sec2}

To address our research questions, whether an underlying emotion co-occurrence network structure which is persistent in time exists and whether this network varies between crisis and non-crisis contexts, we conduct three analyses of the emotion co-occurrence network at different levels of granularity as well as on 7 databases in different periods of crises and non-crises. First, we assess the temporal stability (i.e., persistence) of emotion networks for each dataset, focusing specifically on the rank order of emotion link strengths, measured using Spearman’s rank correlation; Second, we compare network structures across the different datasets, in terms of link rankings, to examine network stability between crisis and non-crisis events. These analyses revealed that the rank of emotion links - as measured by Spearman's rank correlation - remain significantly stable across situations and over time. Despite the stable rank correlation within and between emotions, we observe significant changes in the strength of emotion links during crises compared to the non-crises periods. As such, our third analysis compares the strength of emotion links in crisis datasets with those in non-crisis datasets, highlighting how these strengths vary across different crises. For example, emotion links associated with Tension are found to be strengthened significantly during earthquakes and pre-vaccination period as compared to normal-time, links associated with Fatigue become more pronounced after the start of vaccination (\textbf{Fig. 4}).

We define an emotion as being characterized by several emotion concepts (e.g., cognition, feeling, behavior). We base our analysis on the Profile of Mood States (POMS) method \cite{bib13} — a psychological rating scale—to characterize six emotion dimensions: Tension, Depression, Anger, Vigor, Fatigue, and Confusion. We first extract representative keywords from each of the 65 items in the POMS questionnaire, each questionnaire corresponding to one of the six emotion dimensions. Then, we expand these keywords to construct a comprehensive emotion dictionary containing 792 words (see \textit{Methods}). We consider these words as POMS emotion concepts. For our analysis, we include all posts containing at least one POMS concept (word), regardless of whether they were written by the same users. This approach is equivalent to repeated cross-sectional random sampling \cite{bib62} in psychological research, making it well-suited for capturing emotion prevalence at the population level and offering practical implications for public mental health surveillance and intervention~\cite{bib14}. Although these emotion dimensions contain different numbers of words, in SI–Fig. S7 we applied a sensitivity analysis in which we randomly sample the same number of concepts (words) from each emotion dimension and demonstrate the robustness of our results.

Using this POMS dictionary, we extract co-existing pairs of concepts from individual tweets (\textbf{Fig. 1a}). Then we select the concept links that co-occur significantly more frequently than random combinations of concepts within the same tweet (see \textit{Methods}) to build a POMS concept (word) network. Based on this concept network (\textbf{Fig. 1b}), we construct an emotion network which represents the co-occurrence structure between the six emotions dimensions (\textbf{Fig. 1c}). In this network, each node represents one of the six POMS emotion dimensions, and the weight of each link reflects the strength of the connection between a pair of emotions. This strength is quantified by the number of significant concept links between the two emotion dimensions, normalized by total number of possible links between two emotions and then rescaled by the median value of all emotion links (see \textit{Methods}).

\subsection*{Examining emotion co-occurrence network stability over time during crisis}\label{subsec1}

We begin by investigating the stability of the emotion co-occurrence network structure and how it evolves over time during each crisis event. For each dataset, we generate temporal network snapshots using time resolutions tailored to the nature of each crisis. Earthquakes are sudden and have immediate impacts, so we adopt a daily time resolution to construct and compare temporal networks. In contrast, for the COVID-19 vaccination dataset, we use a quarter-year time resolution, as the pandemic unfolded over years and vaccinations are administered in repeated phases to establish and sustain population immunity.

To compare the similarity between emotion networks in different times, we use the Spearman's rank correlation, $r_s$ (Equation 1), to test if and how much the rank of emotion links changes over time. The visualization of rank of emotion links over time in each dataset are provided in SI Fig. S1. Let \(S_{t_1}\), \(S_{t_2}\)  be the strengths of the set of emotion links in two time windows, which are converted to ranks \(R[S_{t_1}]\), \(R[S_{t_2}]\):
\begin{equation}
r_s(S_{t_1}, S_{t_2}) = \frac{\mathrm{cov}(R[S_{t_1}], R[S_{t_2}])}{\sigma(R[S_{t_1}]) \, \sigma(R[S_{t_2}])}.
\end{equation}
Where \(\mathrm{cov}(\cdot)\) is the covariance and \(\mathrm{\sigma}(\cdot)\)  is the standard deviation.

The evaluation of the stability of emotion networks across $n$ time windows can be regarded as a multiple comparison with sample size equal to the total number of pairwise comparisons: $n \times (n-1)/2$.  To assess the significance of such multiple hypothesis testing while mitigating the risks of false positives, we adjust the $p$-values using the False Discovery Rate (FDR) \cite{bib57}. The FDR calculates the expected ratio of the number of false positive classifications (false discoveries) among all rejected null hypotheses.


In \textbf{Fig. 2}, we show the similarity of temporal emotion and concept networks within the COVID-19 vaccination datasets as an example, in which we split the 2.5 year's data from January 2020 to June 2022 into 10 quarters. We surprisingly find that even though the concept (word) network exhibits considerable variation as is depicted in the low similarity found in \textbf{Fig. 2b}, the structure of the emotion network based on Spearman's correlations remains stable over time, as shown in \textbf{Fig. 2a}.


Specifically, the Spearman rank correlation of emotion networks across different time windows remains consistently high both among the inter and intra-emotion links (the median value of Spearman's {$\rho$} = 0.87, $p < .001$) as well as the inter-emotion links (the median value of Spearman's $\rho$ = 0.75, $p$ = .01). The Spearman coefficient $\rho$ for the combined intra- and inter-emotion links are higher than those for the inter-emotion network alone (two-sample t-test comparing two sets of Spearman coefficients comparing emotion links across time windows-both intra- and inter-emotion links versus inter-emotion links: $t(88)=4.48$, $p<.001$, $d=0.96$), which can be attributed to a more stable intra-emotion connections across all network snapshots (Fig. S6).

In contrast to the stability of emotion networks, the concept networks, where each concept is a node—constructed from concept links that co-occur significantly more frequently than random combinations—exhibit much lower temporal similarity, as measured by the Jaccard Index (SI Equation~S1). The average Jaccard similarity of these concept networks ranges from 0.27 to 0.49 (\textbf{Fig. 2b}). This indicates that only a relatively small fraction of significant word links is consistently shared across time windows. Further details on the significance test of stability of concept links are presented in SI Fig. S3, which shows that only a limited fraction of links (37\% in the vaccination dataset) appear significantly more stable than would be expected by chance.


Based on the color patterns in \textbf{Fig. 2a} (the part below the diagonal shows emotion network including the inter- and intra-emotion links, while the part above shows only inter-emotion links), the heatmaps of emotion network visually cluster into two distinct periods—before and after the start of COVID-19 vaccination in April 2021. The Spearman correlations of emotion networks between the pre-vaccination and vaccination periods reveal a lower similarity compared to similarity within either the pre-vaccination or vaccination period. Specifically, two-sample t-tests comparing the set of Spearman coefficients within the pre-vaccination or vaccination periods to those between the two periods revealed significant differences for both the combined intra- and inter-emotion networks ($t(43) = 7.3$, $p < .001$, $d = 2.2$), as well as for the inter-emotion networks alone ($t(43) = 7.6$, $p < .001$, $d = 2.3$). This shift suggests that Twitter users’ emotion co-occurrence structures are influenced by external events such as the rollout of vaccination.



\begin{figure}[H]
\centering
\includegraphics[width=1.0\textwidth]{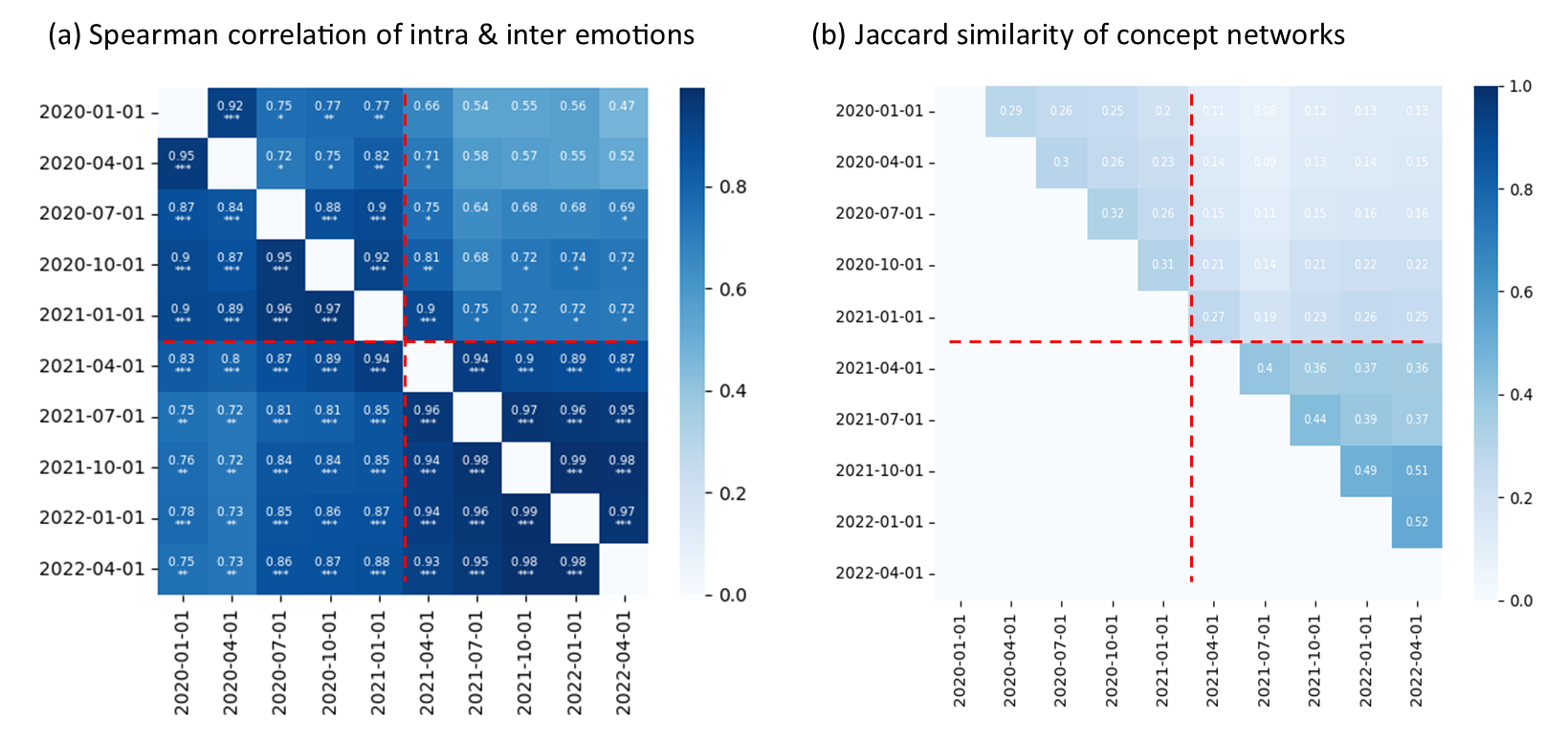}
 \vspace{1mm}
  {\footnotesize \textit{*}: $p \leq$ 0.05,\quad \textit{**}: $p\leq$ 0.01,\quad \textit{***}: $p\leq$ 0.001.}
\caption{Stability of emotion and concept network over time (vaccination dataset). (a) Heatmap of Spearman rank correlation of emotion link's strength between emotion network snapshots. The part BELOW the diagonal shows the similarity of emotion networks including both intra- and inter-emotion links, while the part ABOVE the diagonal shows that of only inter-emotion links. The annotations indicate Spearman's rank correlation coefficient, along with the significance asterisks which depict the level of significance based on the $p$-value. The $p$-values are evaluated by the FDR. (b) Heatmap of Jaccard Index between temporal concept networks. The annotations indicate the value of Jaccard Index. In all sub-figures, we adopt thresholds of link weight $\ge$ top 10\% and link strength $\ge$ 3 (see \textit{Methods}). The color bar on the right of each subplot shows the mapping between colors and the corresponding data values, ranging from 0 to 1. The red dotted lines represent the commencement of vaccination in Japan in April 2021.}
\end{figure}

We also apply the same method to analyze the similarity of temporal emotion and concept networks for all datasets. \textbf{Table} 1 summarizes the Spearman correlation coefficients and corresponding $p$-values for temporal comparisons of emotion networks within each crisis dataset. The reported values represent the mean $\pm$ standard deviation across all pairs of time windows. Corresponding heatmaps illustrating network similarity over time are shown in SI Fig. S2. The detailed analysis on the stability of concept word networks over time for all datasets are presented in SI Fig. S3 to Fig. S5 and Table S1. 

We observe a consistent persistence pattern across all datasets: the emotion networks remain relatively stable over time within each situation as measured by Spearman's rank correlations (see \textbf{Table~1} for detailed statistics and SI Fig. S1 for visualization of ranks of inter-emotion links over time for each dataset), despite the fact that the underlying concept networks used to construct them exhibit considerable temporal instability (see SI Table~S1). 

For completeness, in addition to comparing the ranks of emotion links, we also evaluated the stability of emotion networks by taking the strength of emotion links into consideration. Figures comparing the strength of emotion links between time windows within each dataset SI Fig. S6), together with the results of Pearson’s correlation (SI Fig. S7) comparing the correlation of emotion link's strength between time windows. These results further demonstrate a high persistence of emotion networks across time within each dataset, even when the emotion link's strength is taken into consideration, consistent with the rank-based findings.



\begin{table}[h]
\caption{Summary of Spearman correlation within each datasets}\label{tab1}%

\begin{tabular}{p{0.5\linewidth} p{0.25\linewidth} p{0.25\linewidth}  }

\hline
    & Intra \& inter emotions: &  Inter-emotions:   \\

    & Spearman coefficient  & Spearman coefficient   \\
\hline
  Pre-vaccine \& vaccination (2020.1 - 2022.6) &  0.87 $\pm$ 0.08 *** & 0.76 $\pm$ 0.14 * \\
\hline
  Pre-vaccination (2020.1-2021.3) &  0.91 $\pm$ 0.04 *** & 0.82 $\pm$ 0.07 ** \\
\hline
  Vaccination (2021.4-2022.6) &  0.96 $\pm$ 0.02 *** & 0.94 $\pm$ 0.04 *** \\
\hline
  Non-earthquake tweets (2011.3.11-3.17) &  0.95 $\pm$ 0.031 *** & 0.90 $\pm$ 0.06 *** \\
 \hline
 Earthquake tweets (2011.3.11-3.17) &  0.95 $\pm$ 0.03 ***  & 0.94 $\pm$ 0.02 *** \\
 \hline
 Non-earthquake tweets (2018.6.18-6.21) &  0.99 $\pm$ 0.01 *** & 0.97 $\pm$ 0.02 *** \\
 \hline
 Earthquake tweets (2018.6.18-6.21) &  0.92 $\pm$ 0.02 *** & 0.90 $\pm$ 0.06 *** \\
 \hline
 Bluesky (2024.6 and 2024.12) &  0.95 $\pm$ 0.03 *** & 0.90 $\pm$ 0.07 *** \\
 \hline
\end{tabular}
{\footnotesize \textit{*}: $p \leq$ 0.05,\quad \textit{**}: $p \leq$ 0.01,\quad \textit{***}: $p \leq$ 0.001.}
\end{table}


\newpage

\subsection*{Comparing emotion co-occurrence structure stability during crises and normal time}\label{subsec1}

Similarly, we calculate the Spearman correlation of emotion networks and Jaccard Index of concept networks across different datasets. \textbf{Fig. 3} shows a comparative analysis of emotion and concept networks across all seven datasets. Specifically, to assess the similarity of emotion networks between two datasets—comprising $n_i$ and $n_j$ temporal network snapshots, respectively—we compute the Spearman rank correlation for each of the $n_i \times n_j$ pairs of emotion networks between the two datasets and take the average correlation as the overall similarity measure see details of average $\pm$ standard deviation in SI Table S2. To account for the multiple comparisons involved in this process, we adjust the $p$-values using the False Discovery Rate (FDR) \cite{bib65}. To assess the similarity of concept networks, we calculate the average Jaccard Index between all temporal concept networks in the pairs of datasets. The significance test on stability of concept links across datasets is presented in SI Fig. S5.



\begin{figure}[h]
\centering
\includegraphics[width=1.0\textwidth]{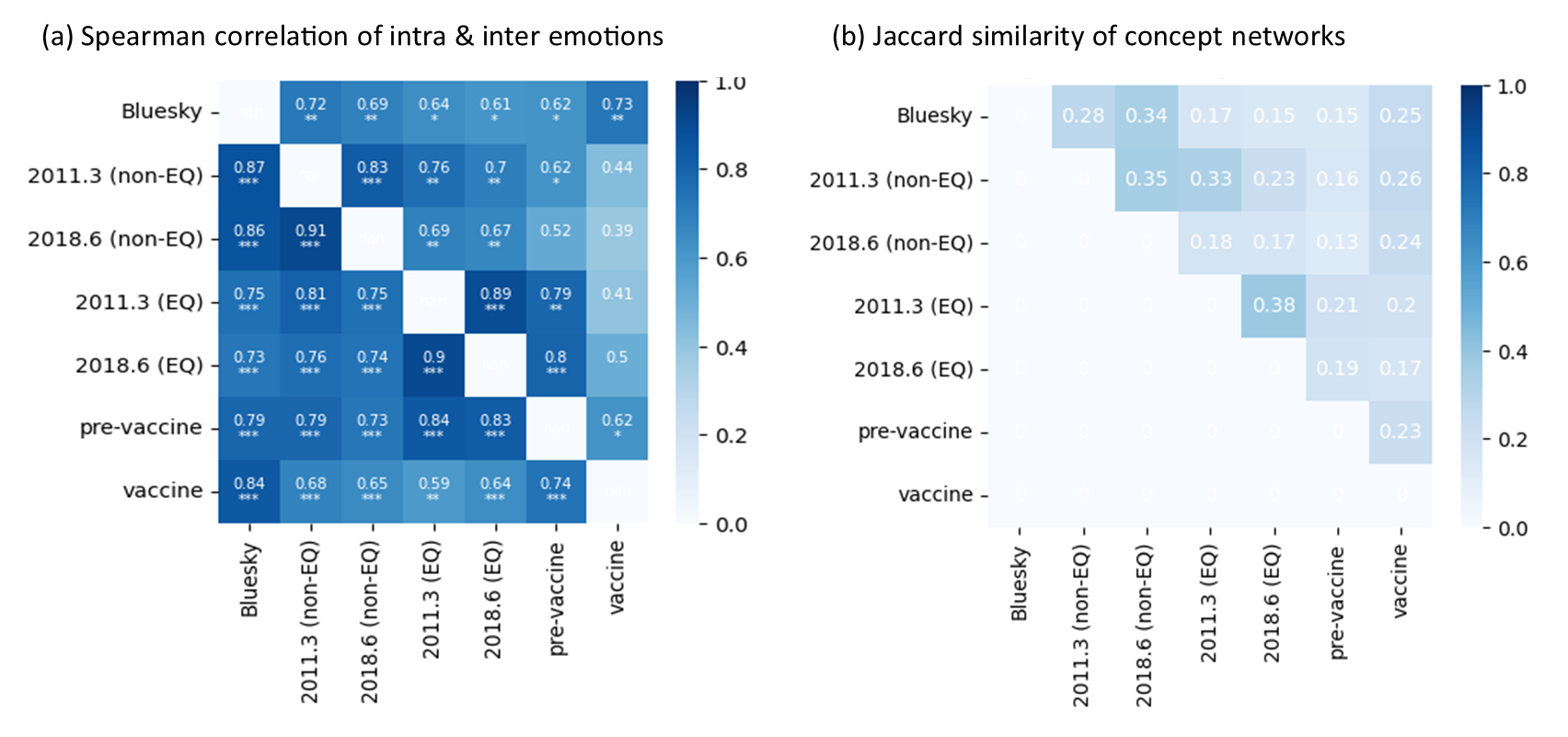}
{\footnotesize \textit{*}: $p \leq$ 0.05,\quad \textit{**}: $p \leq$ 0.01,\quad \textit{***}: $p \leq$ 0.001.}
\caption{Similarity of emotion and concept networks across datasets. (a) Heatmap of Spearman rank correlation of emotion link's strength between emotion networks of different datasets. The part
BELOW the diagonal shows the similarity of emotion networks including both intra- and inter- emotion links, while the part ABOVE the diagonal shows that of only inter-emotion links. The annotations indicate the average Spearman's rank correlation coefficient, along with the significance asterisks which depict the level of significance based on the average $p$-value. The $p$-values are adjusted by the FDR. (b) Heatmap of Jaccard Index between concept networks of different datasets. The annotations indicate the average Jaccard Index. In all sub-figures, we adopt thresholds of link weight $\ge$ top 10\% and link strength $\ge$ 3 (see \textit{Methods}). The color bar on the right of each subplot shows the mapping between colors and the corresponding data values, ranging from 0 to 1. }\label{fig1}
\end{figure}


Interestingly, we find that the structure of emotion networks remains consistently stable across datasets, even though the underlying concept networks used to construct them exhibit a much higher variability. Specifically, the intra- and intra-emotion networks (\textbf{Fig.~3a} below diagonal) demonstrate strong consistency, with an average Spearman correlation of $\rho = 0.87$ ($p < 0.001$), while the inter-emotion networks (\textbf{Fig.~3a} above diagonal) also show considerable stability, with an average Spearman correlation of $\rho = 0.76$ ($p < 0.001$). In contrast, the concept networks exhibit significantly low stability: the average Jaccard Index across time windows is only 0.23 (\textbf{Fig.~3b}) and just 39\% of concept links show significant stability. Here, stability of concept links is defined as the probability that a given concept link is repeatedly significant (weight $\ge$ top 10\% and strength $\ge$ 3) across time at a frequency higher than expected under random resampling (where the same number of significant links is drawn from all possible links; see SI Fig. S4 and Fig. S5 for details).

In the following, we focus on a more detailed comparison of the emotion networks across different situations. We begin by comparing the emotion networks observed during crisis and non-crisis events. The top row of \textbf{Fig. 3a} presents the Spearman correlations between the emotion networks derived from all datasets collected during crises and the Bluesky dataset (the non-crisis periods). Notably, the emotion networks during crisis periods exhibit significant similarity to those of normal times, suggesting that the underlying emotion co-occurrence structure in normal times remains largely stable during crises. These datasets span a period of over 13 years, suggesting that the emotion co-occurrence structure, as reflected in the rank order of emotion links, exhibits long-term stability.

Notably, as is depicted in \textbf{Fig. 3a}, the rank of inter-emotion links in the Bluesky dataset (normal-time) show a more significant similarity with the non-earthquake related datasets (during the 2011 earthquake - Spearman's $\rho = 0.72$, $p = .003$; during the 2018 earthquake - $\rho = 0.69$, $p = .004$), as compared to the earthquake-related datasets (during the 2011 earthquake $\rho = 0.64$, $p = .021$; during the 2018 earthquake $\rho = 0.61$, $p = .023$). Moreover, the similarity between normal-time and vaccination dataset ($\rho = 0.73$, $p = .002$) is more significant than pre-vaccination dataset ($\rho = 0.62$, $p = .03$). Likewise, when comparing the emotion network between crises, we find a higher similarity between earthquake-related datasets and pre-vaccination datasets (2011 earthquake vs. 2018 earthquake - $\rho = 0.73$, $p < .001$; 2011 earthquake vs. pre-vaccination - $\rho = 0.79$, $p = .001$; 2018 earthquake vs. pre-vaccination - $\rho = 0.8$, $p < .001$), than when comparing to the vaccination dataset (2011 earthquake vs. vaccination $\rho = 0.44$, $p = .21$, 2018 earthquake vs. vaccination $\rho = 0.5$, $p = .06$, pre-vaccination vs. vaccination $\rho = 0.62$, $p = .02$). These results suggest that during earthquakes and pre-vaccination periods, when the crises are heightened and comparable, user’s emotions associated with Tension are amplified. In contrast, after the roll-out of COVID-19 vaccines, likely due to their effectiveness in mitigating the pandemic, the emotion network structure relaxes and reverted toward a baseline-the normal-time structure.

These reported results are based on thresholds (weight $W \geq 10\%$ and strength $S \geq 3$) used to filter significant concept links. In SI Fig. S10 and Fig. S11, we show that also when the thresholds are kept above weight $W \geq 30\%$ and strength $S \geq 1$—which excludes a number of low-frequency and potentially random links—the results remain robust.

\subsection*{Comparing the strength of emotion links between crisis and non-crisis events}\label{subsec1}

So far, we observed a stable emotion co-occurrence structure—as measured by the Spearman rank correlation—both over time within specific situations and across different situations. However, even though the order of emotion links remain stable, their strength may vary based on situations. Elucidating such link strength variability can highlight how collective emotional responses are shaped by crises which could help in better design of targeted mental health interventions. To examine in detail the strength of specific emotion links and how they vary across contexts, we represent the emotion networks for each dataset and compares the strength of emotion links across them (\textbf{Fig. 4}).


Because the strength of an emotion link is calculated based on the number of significant concept (word) links between emotions (see \textit{Methods}), a dataset with a smaller volume or spans over a short period tends to have less concept links which leads to a lower emotion strength, making it difficult to compare the strength of emotion links across datasets and across time. To address this issue, we rescale the emotion links by assuming that the median levels of emotion links are consistent within each network snapshot. Specifically, for each temporal emotion network including both inter- and intra-emotions, we normalize the strength of each emotion link over the median value of the all emotion links to calculate the rescaled emotion strength. The temporal rescaled strength of emotion links for each dataset is visualized in SI Fig. S8.

In \textbf{Fig. 4a}, we compare the respective emotion links' strengths across datasets. Each line in the line chart on the top of \textbf{Fig. 4a} represents the rescaled emotion strength for a single dataset, showing the mean along with the 75\% confidence interval. The black line corresponds to the Bluesky dataset (normal time), which serves as the baseline for our comparison. Colored lines represent emotion strength during crisis periods. Overall, intra-emotion links—those connecting the same emotional dimensions—are significantly stronger than inter-emotion links (two-sample t-test comparing the set of intra-emotion links with inter-emotion links based on all emotion network snapshots: $t(880) = 19.1$, $p < .001$, $d = 1.4$), particularly the Fatigue–Fatigue, Tension–Tension, and Depression–Depression links as is shown in \textbf{Fig. 4a}. 
To better visualize the inter-emotion links in greater detail, we present an inset figure displaying the log-transformed strengths of these links. We can observe that the inter-emotion links that are generally strong across all datasets are Fatigue-Tension. In contrast, Vigor-Confusion, Anger-Vigor and Depression-Confusion tend to be mostly low in strength.

To assess how significantly each emotion link deviates from the baseline, we perform a two-sample t-test comparing the rescaled emotion link strengths between datasets - ${S'{ij}^{(t)}}\text{dataset1}$ and ${S'{ij}^{(t)}}\text{dataset2}$, where $i$ and $j$ represent two emotions, and $t$ represents different time windows within each dataset. In the heatmap at the bottom of \textbf{Fig. 4a}, we visualize the t-test statistics, with red depicting strengthened emotion link strength, and blue depicting weakened strength. The asterisks represent the significance of the t-test. Details of the t-test statistics are provided in SI Table S3.

Comparing the crisis datasets to Bluesky dataset (normal time), we find that the strength of most emotion links changes significantly (SI Table S3), even though the ranking order of emotion links remains significantly stable, as we explained earlier. The relatively stable emotion links are Anger-Fatigue, Anger-Tension and Depression-Confusion.

We observe that the two non-earthquake related datasets collected during earthquake periods show a different deviation from the earthquake-related datasets and pre-vaccination datasets. This observation aligns with the previous section where we compare the rank of emotion links across datasets. Specifically (detailed t-test statistics are provided in SI Table S3), in the non-earthquake related datasets, we observe that the emotion links that significantly strengthened as compared to normal-time are mostly Anger-related: Anger-Depression, Anger-Vigor, Anger-Confusion, and Depression-Confusion; while emotion links that significantly weakened are mostly Fatigue related: Fatigue-Fatigue, Fatigue-Vigor, and Fatigue-Confusion. 

In contrast, in the earthquake-related datasets and pre-vaccination dataset, we can see more emotion links that significantly changed as compared to normal-time. The emotion links that significantly strengthened are mostly related to Tension, indicating heightened expressions of stress and anxiety: Tension-Tension, Tension-Vigor, Tension-Confusion, and Depression-Fatigue; while emotion links that significantly weakened are: Anger-Anger, Depression-Depression, Anger-Depression, Fatigue-Confusion, and Vigor-Confusion.

Notably, after the COVID-19 vaccination started, the strength of emotion links changed differently from that of the pre-vaccination period. Specifically, the links that are associated with Fatigue and Confusion, that is, Fatigue-Fatigue, Confusion-Confusion, Fatigue-Vigor, Fatigue-Tension, Fatigue-Confusion, Vigor-Tension, and Tension-Confusion,  have been strengthened significantly likely reflecting increased discussions around physical side effects such as tiredness and inertia.

To better visualize the comparison of relative strength between inter-emotion links and across datasets, we show \textbf{Fig. 4b} which depicts the emotion networks across all datasets, with a node representing each emotion, and the thickness of the links represents the average strength of the rescaled emotion links $\bar{S'}_{ij}$. 

As can be observed visually (also supported by statistic comparison showing in \textbf{Fig. 4a}), the datasets of non-earthquake related posts during the 2011 earthquake (\textbf{Fig. 4b2}) and the 2018 earthquake (\textbf{Fig. 4b4}), show smaller deviations from normal times compared to the datasets filtered using earthquake-related keywords (\textbf{Fig. 4b3} and \textbf{Fig. 4b5}.) Among the emotion links, those most strongly strengthen in the earthquake-related datasets are Fatigue–Tension ($t(17) = 3.59$, $p = .002$, $d = 1.75$), followed by Depression–Fatigue ($t(17) = 4.11$, $p < .001$, $d = 2.01$) and Vigor–Tension ($t(17) = 5.54$, $p < .001$, $d = 2.7$). These shifts likely reflect emotional responses specific to the acute stress and uncertainty of natural disasters. When comparing the pre-vaccination (\textbf{Fig. 4b6}) and vaccination (\textbf{Fig. 4b7}) datasets, we find that after the start of the vaccination period, a notable strengthening of Fatigue-related emotion links can be observed, including Fatigue–Tension, Fatigue–Vigor and Fatigue–Confusion, with all $p$'s $< .001$ as is shown in Table S3. These changes may reflect complicated emotional responses to the physical side effects associated with the vaccination process.



\begin{figure}[H]
\centering
\includegraphics[width=0.85\textwidth]{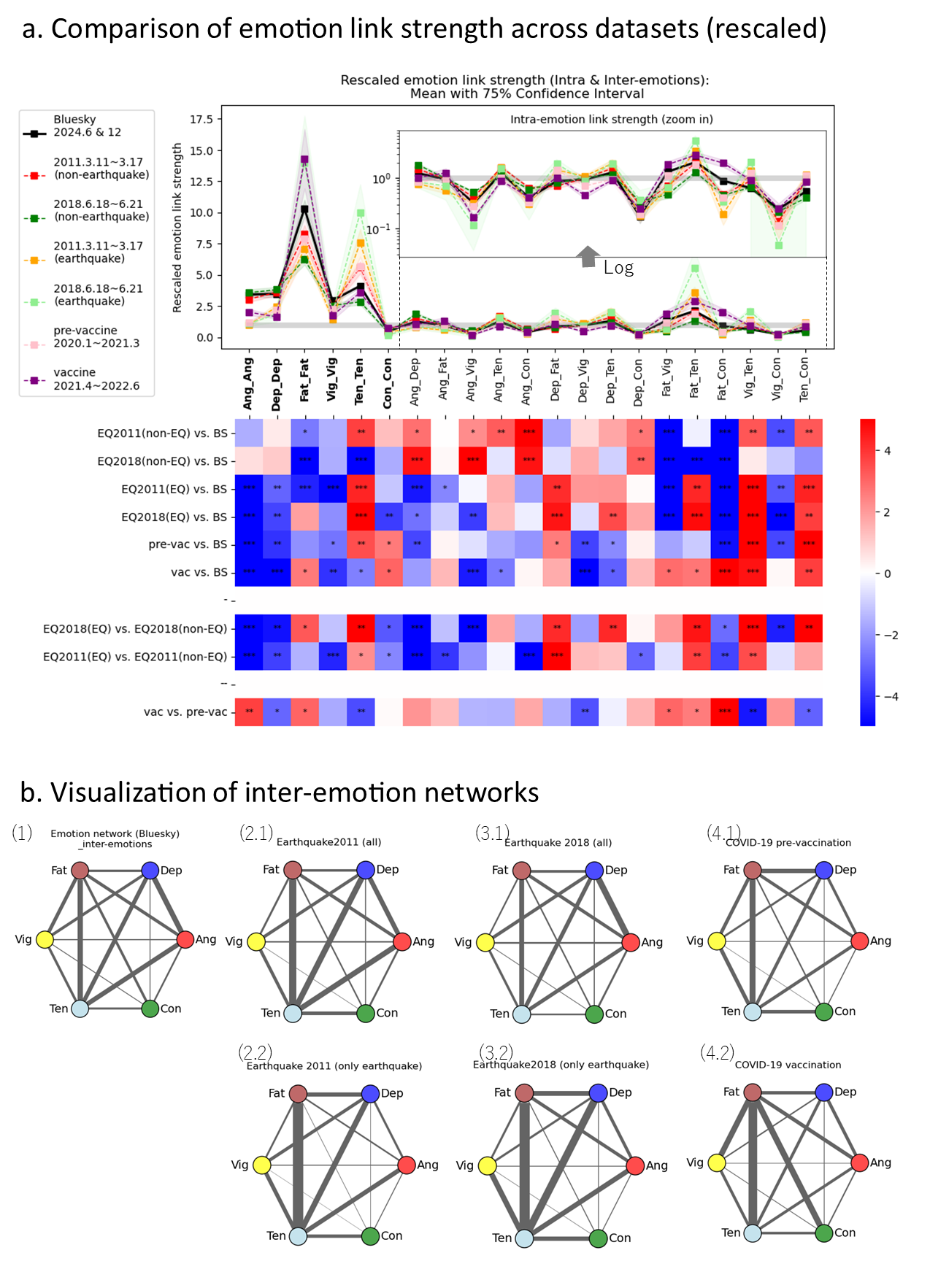}
\caption{Comparing between the strength of emotion links in different datasets. (a) Visualization of emotion link strengths and the two-sample t-test across datasets. The strength of emotion links is rescaled so that the median of the strengths of each dataset becomes 1. The line charts show the median strength of inter- and intra-emotion links with 75\% confidence interval for each dataset. The inset figure zooms into the logarithmic strength of inter-emotion links to better visualize the comparison between different datasets. The heatmap shows the statistics of two-sample t-tests to compare the distribution of emotion link strength between different datasets, with red color representing an increment of strength and blue color representing a reduction of strength. The asterisks show the significance of such changes, measured by the $p$-value of the $t$-test - \textit{*}: $p \leq$ .05,\quad \textit{**}: $p \leq$ .01,\quad \textit{***}: $p \leq$ .001. (b) Visualization of inter-emotion networks. The node represents emotion dimensions, thickness of links depicts the rescaled strength of link between two emotions.
Abbreviation of datasets: EQ2011(All) - Earthquake2011 (all posts), EQ2011- Earthquake2011 (filtered by earthquake-related keywords), EQ2018(All) - Earthquake2018 (all posts), EQ2018- Earthquake2018 (filtered by earthquake-related keywords), BS: Bluesky.}\label{fig1}
\end{figure}

\section{Discussions}\label{sec4}

In this study, we introduce a novel approach to quantifying emotion co-occurrence by creating an emotion network based on identifying statistically significant associations between emotion concepts (words). Using large-scale Japanese social media datasets spanning 13 years, we examine the emotion network structure based on co-occurrence of emotion concepts within individual posts across both crisis and normal times.

Our findings reveal a stable emotion co-occurrence within individual social media users in Japan. Despite the dynamic variation nature of word usage, the underlying emotion network based on co-occurrence structure remains remarkably stable both in the short term (daily) and over longer time scales (quarters and years). In the emotion network we find that among the co-occurring emotion links, the intra-emotion links are generally stronger than the inter-emotion links, especially the Fatigue-Fatigue, Tension-Tension, and Depression-Depression links. The inter-emotion link that is relatively stronger are Fatigue-Tension. We also observe some emotion links that are consistenly low in all time periods, including Vigor-Confusion, Anger-Vigor and Depression-Confusion.

 Despite the general perspective in psychological literature that emotion co-occurrence is highly situation-dependent space \cite{bib10,bib11,bib47}, in our study based on large-scale datasets including a large population, we find that the fundamental structure of emotion co-occurrence, as represented by the rank of emotion links, is generally stable and largely preserved across both crisis and non-crisis periods. Interestingly, when comparing Spearman correlations between datasets, the non-earthquake-related datasets collected during earthquake periods show greater similarity to normal-time data than the earthquake-related datasets. By contrast, the earthquake-related datasets and the pre-vaccination datasets, both reflecting intense crisis situations, exhibit high similarity with each other. Furthermore, the post-vaccination emotion network appears more similar to the normal-time network than to the pre-vaccination one. 

 Although the rank of emotion links remains stable across datasets and over time, the strength of these links exhibits significant changes during crisis periods compared to normal times. For example, the earthquake-related datasets and the pre-vaccination dataset show more significant shifts of emotion links than the non-earthquake-related datasets, with a particularly strong increase in links associated with Tension. During the COVID-19 vaccination period, links involving Fatigue become stronger, likely reflecting public discourse around vaccine side effects.

Another interesting finding of our research is that the emotion network structure based on the Bluesky dataset (representing non-crisis event) exhibits a high degree of similarity to that of the Twitter dataset. Recent studies have shown that after X (formerly Twitter) implemented new moderation policies, many users with differing ideological views migrated to alternative platforms such as Bluesky, which tend to display significantly higher levels of ideological homogeneity \cite{bib38,bib39}. Despite the polarized political and ideological environments of Twitter and Bluesky, our research finds that the underlying emotion network structures remain consistent across both platforms.

Furthermore, comparing our results with previous research \cite{bib9,bib45,bib10}, we observe common patterns in the structure of emotion co-occurrence. Lange et al. \cite{bib9} constructed a network between emotion concepts based on surveys and found that the within-emotion links are stronger than the between-emotion edges, which is aligned to our findings (see \textbf{Fig. 4a}). Vansteelandt et al. \cite{bib45} studied the co-occurrence of emotions within individuals in daily life using experience sampling methodology, and found a positive correlation between emotions links-fear and anger (0.6), sadness and anger (0.38), sadness and fear (0.42). In our study, we observe a similar strong co-occurrence of these emotion links during normal time (Tension-Anger, Depression-Tension and Anger-Depression, as is depicted by the black line in \textbf{Fig. 4a}). Moeller et al. \cite{bib10} found that the most frequently co-occurring emotions among high school students are tired–stressed, tired–bored, and tired–happy. Interestingly, we observe similar connections in our data—specifically, Fatigue–Tension, Fatigue–Confusion, and Fatigue–Vigor—with particularly strong co-occurrence during the COVID-19 vaccination period (see \textbf{Fig. 4b7}).

Our approach can be extended to other emotion rating scales, such as PANAS \cite{bib37}, Plutchik’s Wheel of Emotions \cite{bib36}, and the Depression Anxiety and Stress Scales (DASS) \cite{bib35}, all of which define emotion dimensions consisting of multiple sub-concepts. In terms of data sources, our method is applicable to a wide range of textual data, including social media posts, online forum discussions, customer reviews, and open-ended survey responses, enabling broad and scalable emotion analysis across various contexts. Moreover, it offers a systematic framework for analyzing and comparing emotion co-occurrence networks across different datasets, facilitating cross-domain and longitudinal emotional research. 


We acknowledge that several limitations should be considered. First, the large-scale social media datasets used in this study do not include user identity or demographic information, which limits our ability to examine individual differences or subgroup behaviors. Second, our approach is based on cross-sectional sampling, meaning we do not track the same users over time; hence, our findings reflect collective patterns rather than individual emotional dynamics. Third, our study focuses solely on Japanese-language posts, and the results may not generalize to other cultural or linguistic contexts. Fourth, our research is primarily observational rather than inferential; that is, we analyze the co-occurrence structure of emotions based on empirical data, but do not attempt to causally explain why certain emotion links are stronger than others. Investigating the underlying causal mechanisms remains an important direction for future work. Finally, emotion detection is conducted using a dictionary-based method derived from the POMS scale. While this provides interpretability and standardization, recent advancements in large language models (LLMs) offer improved accuracy and could be explored in future research.

To the best of our knowledge, this study is the first to examine the stability of the structure of emotion network based on co-occurrence among social media users and how it changes during crises, using data driven computational approaches utilizing large-scale datasets. Our findings contribute not only to a deeper understanding of the complexity of emotions across both normal and crisis periods at the population level, but also offer a novel approach for studying multiple emotions and their relations, which could move forward the theoretical understanding of the complexity of emotions. This approach can be utilized by researchers, mental health professionals, and policymakers to monitor public sentiment, detect emotional disruptions during crises, and inform timely and targeted interventions.

\section{Methods}\label{sec3}

\subsection*{Dataset}\label{subsec3}

To study the emotion co-occurrence under different situations and over short and long-term, we analyze four different datasets containing social media posts written in Japanese: Twitter data during 2011.3.11 Tohoku earthquake, Twitter data during 2018.6.18 Osaka earthquake, Twitter data before and during COVID-19 vaccination and Bluesky data in June and December 2024 that are considered to be normal time. In total, these dataset sums up to 48 million tweets that contains at least one POMS word.

Both the 2011 and 2018 Osaka earthquake datasets contain tweets posted by users all over Japan, who may not be directly impacted by the earthquake. Therefore, to further zoom into tweets directedly associated with earthquake, we filter by earthquake-related keywords and create two additional datasets.

The vaccination dataset consists of tweets containing the keyword "vaccine" posted during 2.5 years since January 2020 when the COVID-19 pandemic broke out till May 2022 when the third dose of vaccination completed. In Japan, the first and second doses of COVID-19 vaccination started in April 2021 first for the elderly and medical workers, then extended to the public in July 2021. We hypothesize that the emotion reaction would be different before and after vaccination started in April 2021, and therefore split the dataset into two parts. As such, we have a total of seven datasets. \textbf{Table} 1 provides the details of these datasets.

\begin{landscape}
\begin{table}[ht]
\centering
\caption{Overview of the datasets used in this study}
\begin{tabular}
{|p{0.12\linewidth} | p{0.13\linewidth}  | p{0.1\linewidth}  | p{0.1\linewidth} | p{0.1\linewidth} | p{0.1\linewidth}  | p{0.1\linewidth} | p{0.1\linewidth}|}
\hline
\textbf{Platform} & \textbf{Twitter\newline 2011 Tohoku earthquake\newline (non-earthquake tweets)} & \textbf{Twitter\newline 2011 Tohoku earthquake\newline (only earthquake)} & \textbf{Twitter\newline 2018 Osaka earthquake\newline (non-earthquake tweets)} & \textbf{Twitter\newline 2018 Osaka earthquake\newline (only earthquake)} & \textbf{Twitter\newline COVID-19\newline pre-vaccination} & \textbf{Twitter\newline COVID-19\newline vaccination} & \textbf{Bluesky\newline Normal time} \\
\hline
\textbf{Period} & 2011.3.11– 2011.3.17 & 2011.3.11– 2011.3.17 & 2018.6.18– 2018.6.21 & 2018.6.18– 2018.6.21 & 2020.1– 2021.3 & 2021.4– 2022.5 & 2024.6 and 2024.12 \\
\hline
\textbf{Description} & Tweets excluding ``earthquake'' related keywords & Tweets containing ``earthquake'' related keywords & Tweets excluding ``earthquake'' related keywords & Tweets containing ``earthquake'' related keywords & Tweets posted in Japanese containing the keyword ``vaccine'' & Tweets posted in Japanese containing the keyword ``vaccine'' & All tweets posted in Japanese \\
\hline
\textbf{Tweets with at least one POMS word} & 14 million & 3 million & 15 million & 1.6 million & 2 million & 7 million & 6 million \\
\hline
\textbf{Time windows} & Daily (8 time windows, splitting the first day into two time windows, before and after the earthquake) & 7 days & Daily (5 time windows, splitting the first day into two time windows, before and after the earthquake) & 4 days & 5 quarters & 5 quarters & 8 weeks  \\
\hline
\end{tabular}
\end{table}
\end{landscape}

\subsection*{Emotion dictionary based on POMS questionnaire}\label{subsec4}

For extracting emotion concepts from social media posts, we build an emotion dictionary by adopting emotion categories introduced in POMS (Profile of Mood States), which is a questionnaire-based psychological rating system developed by McNair, Droppleman, and Lorr in 1971 \cite{bib13}. It consists of six emotion dimensions: Tension, Depression, Anger, Vigor, Fatigue and Confusion. Each emotion dimension has a few emotion concepts to describe the different perspectives of emotions. 

POMS rating scale has been widely adopted by psychologists and psychotherapists, for example, for measuring athlete's mood states before and after competitions \cite{bib15}, and for tracking mental health of patients with chronical diseases \cite{bib16}. The traditional way of using this emotion scale is giving each participant a questionnaire consist of 65 questions, each represents an emotion concept (e.g. worn out, unhappy, confused) and is categorized under one of the six emotion dimensions. A participant is required to respond to each question by rating his/her feelings at the scene with a scale from 1 (not at all) to 5 (extremely).  Recently, researchers extracted the keywords in each question and the corresponding emotion dimension in POMS to build a dictionary for identifying emotions in social media data \cite{bib4,bib5,bib17}. 

In our study, we build an expanded emotion dictionary based on the Japanese version of the POMS questionnaire. Researchers have pointed out the drawbacks of the traditional psychological rating scale, that it may fail to capture the granular emotions that are not provided in the questionnaire \cite{bib19,bib20}. Furthermore, the informal and spontaneous language commonly used on social media presents additional challenges for mapping users' feelings to predefined categories in the POMS scale.

We use a pre-trained Word2Vec (word embedding) database \cite{bib18} that maps each Japanese word into a high-dimensional space to capture the relationships between words. In the database, words that appear in a similar context (for example, synonymous words or words with the same meaning but are spelled differently) tend to be close in cosine similarity distance. For each POMS word, we select the surrounding words from the Word2Vec database that have a cosine similarity of more than 0.7, then we manually screening through the words to make sure that they are relevant to the respective emotion. Finally, we align the spelling for the same words (for example, the same words with different tenses) to ensure that each word in the dictionary is independent. In such a way, we build a POMS dictionary that consists of 792 words under six dimensions: Anger (139 words), Confusion (197 words), Depression (109 words), Fatigue (70 words), Tension (121 words) and Vigor (156 words). The dictionary is provided in the SI.

\subsection*{Building a concept (word) network with significant links}\label{subsec5}

In recent years, there has been growing interest in applying network theory to psychological research to investigate the complex associations among psychological concepts. For example, Moeller et al.\ surveyed high school students about their feelings toward school life and constructed networks connecting co-occurring emotions within individual participants~\cite{bib10}. Kenett et al.\ employed a free association generation task, using network theory to model the associative relationships between target words and to explore semantic differences between individuals with high and low levels of creativity \cite{bib67}. Similarly, Beck et al.\ examined the associations among personality concepts by constructing networks in which links represented partial correlations between pairwise time series of personality processes within individuals~\cite{bib33}.


In our research, we assume that if a user uses multiple emotion-related words in a given post, then these emotion concepts co-occur within this person. We create an undirected and weighted network to represent such a co-occurring relationship. A node represents a word describing an emotion concept, and a link connecting a pair of words represents the emotion concepts' co-occurrence within the same post. By aggregating over multiple tweets, we can calculate the weight for each concept link, which is represented by the number of posts which this concept link occurs.

However, such weight calculated from word co-occurrence frequency are biased by the frequency of words, that is, a frequently used word tend to have a higher frequency to connect with other words. To test how significant a concept link as compared to a random case, we make the following hypothesis. Let \(l_{w_i,w_j}\) and \(l'_{w_i, w_j}\)  be the actual and random weights between words \(w_i\) and \(w_j\).

\begin{itemize}
  \item 	\textbf{Null hypothesis}: A concept link's occurrence is less than or similar to a random connection, \( l_{w_i, w_j} \leq l'_{w_i, w_j} \)
  \item 	\textbf{Alternative hypothesis}: A concept link's occurrence is significantly larger than a random connection, \( l_{w_i, w_j} >> l'_{w_i, w_j} \)
\end{itemize}

To create a random concept network, we permute the connections between words randomly while keeping the following two conditions unchanged: (1) frequency of each POMS word, (2) number of POMS word per tweet. Specifically, as is depicted in SI Fig. S13a, we list down the post ID that contains at least one POMS word under the first column and the POMS words that appear under the respective post ID under the second column. Then we shuffle the POMS words in the second column randomly until there are no duplicate POMS word in the same post. As such, the POMS words are randomly connected while the two conditions are met. We repeat this reshuffling process for a hundred times, and for each concept link we can get a set of random link weights \(\left\{ l'_{w_i, w_j} \right\}\). In Equation 3, we calculate the strength of a concept link \(w_i\),\(w_j\)  based on the z-score that compares the actual link weight \( l_{w_i, w_j}\)  to the set of random link weight  \(\left\{ l'_{w_i, w_j} \right\}\).  \(\left\langle l'_{w_i, w_j} \right\rangle\) represents the ensemble average of the random weight and \(\sigma\left( l'_{w_i, w_j} \right)\) refers to the standard deviation. SI Fig. S13b gives two examples of concept links to illustrate the distribution of random weights and compare it to the actual weight. 


\begin{equation}
\mathrm{Strength}(w_i, w_j) = \frac{l_{w_i, w_j} - \left\langle l'_{w_i, w_j} \right\rangle}{\sigma\left( l'_{w_i, w_j} \right)}
\end{equation}


However, a link with a high strength does not necessarily mean that it is significant. For example, a concept link that occurs only once will likely have 0 occurrence when randomly permuted, resulting in a high z-score. Therefore, we select the significant links based on two thresholds: strength threshold \(S\) (e.g. \( S \geq 3 \)
, that is, larger than random weights' mean +3 standard deviation) and weight threshold W (e.g. \(W \geq\) the top 10\% of weights). In SI Fig. S14, we show a scatterplot of the concept links in the vaccine dataset. Each node represents a concept link, and its weight and strength are represented by x-axis and y-axis, respectively. The blue nodes above the thresholds (\(S \geq 3\), \(W \geq 10\%\) which is equivalent to concept link occurrence above 15) are the significant links that we want to focus on. In SI Fig. S14a, we provide a visualization of an example concept network with significant links.

The significance of a concept link can fluctuate over time, even when its occurrence remains consistently high. In SI Fig. S14b, we present two examples of frequently occurring concept links from the vaccination dataset—one that consistently maintains high significance and another that exhibits substantial volatility. This variability in significance contributes to the low similarity observed between concept networks constructed across different time windows, as shown in the Results section (\textbf{Fig. 2a}).

\subsection*{Building an emotion network from significant concept links}\label{subsec5}

We argue that a higher number of shared co-occurring emotion concepts between two emotions indicates a stronger co-occurrence between them. Therefore, we construct an emotion network where each node represents one of the six emotion dimensions, and the weight of links between two emotions depicts the number of significant concept links between two emotions. To account for the unequal number of words in each emotion dimension, we normalize the number of significant concept links by the total number of possible word pairs between the two emotions (Equation 3).

\begin{equation}
s_{ij}^{(t)} = \frac{\left|\mathcal{L}_{ij}^{(t)}\right|}{\left|\mathcal{P}_{ij}\right|}
\end{equation}
where $s_{ij}^{(t)}$ denotes the strength of the emotion link between emotions $i$ and $j$ during time window $t$, $\mathcal{L}_{ij}^{(t)}$ is the set of significant concept links observed between emotions $i$ and $j$ and $\mathcal{P}_{ij}$ is the set of all possible concept links between emotions $i$ and $j$, $|\cdot|$ denotes the cardinality (number of elements) of a set.

Because the emotion strength depends on the activity level of each dataset—datasets with more tweets tend to have more concept links—we rescale the emotion strength $s_{ij}^{(t)}$ by its median value to remove this dependency and enable comparison across datasets (see Equation~4)~\cite{bib30,bib61}.

\begin{equation}
{S'}_{ij}^{(t)} = 
\frac{s_{ij}^{(t)}}{
\operatorname{median}\!\left( 
\left\{ s_{ij}^{(t)}\right\}
\right)}
\end{equation}

where $s_{ij}^{(t)}$ denotes the strength of the emotion link between emotions $i$ and $j$ during time window $t$ of a given dataset, and $\operatorname{median}(\cdot)$ represents the median value of a set of values.

\newpage
\bmhead{Acknowledgements}

S.H. and Y.N.K were partially funded by the European Union under the Horizon Europe grant OMINO-Overcoming Multilevel Information Over-load (grant number 101086321, http://ominoproject.eu). Views and opinions expressed are those of the authors alone and do not necessarily reflect those of the European Union or the European Research Executive Agency. Neither the European Union nor the European Research Executive Agency can be held responsible for them. This work was supported by JSPS KAKENHI Grant Number 23K28192.

\bmhead{Competing Interests}
The authors declare no competing interests.

\bmhead{Author contributions}
M.T., S.H., Y.N.K. and O.L. led the project, designed the overall research plan, and supervised the manuscript preparation. Q.W. analyzed the raw data, performed numerical calculations, and drafted the manuscript. O.L. initiated the research idea and supported the checking and interpretation of the results. Y.N.K. also provided psychological interpretations of the results. Y.S. and H.T. contributed to the development of data analysis methods. All authors reviewed and approved the final version of the manuscript.

\bmhead{Data Availability}
The X (Twitter) and Bluesky data cannot be open to the public due to privacy policy, but similar data can be obtained using X (Twitter) API (https://developer.twitter.com/en/products/twitter-api) and Bluesky API (https://docs.bsky.app/). However, aggregated and anonymized data can be shared upon request. For data inquiries, please contact the corresponding author Misako Takayasu (takayasu@comp.isct.ac.jp).







\newpage


\newpage

\bibliography{sn-bibliography}

\begin{thebibliography}{10}
\expandafter\ifx\csname url\endcsname\relax
  \def\url#1{\burl{#1}}\fi
\expandafter\ifx\csname urlprefix\endcsname\relax\def\urlprefix{URL }\fi
\providecommand{\bibinfo}[2]{#2}
\providecommand{\eprint}[2][]{\url{#2}}
\providecommand{\doi}[1]{\url{https://doi.org/#1}}
\bibcommenthead

\bibitem{bib1}
\bibinfo{author}{Abbas, J.}, \bibinfo{author}{Wang, D.}, \bibinfo{author}{Su, Z.} \& \bibinfo{author}{Ziapour, A.}
\newblock \bibinfo{title}{The role of social media in the advent of covid-19 pandemic: Crisis management, mental health challenges and implications}.
\newblock \emph{\bibinfo{journal}{Risk Management and Healthcare Policy}} \textbf{\bibinfo{volume}{Volume 14}}, \bibinfo{pages}{1917--1932} (\bibinfo{year}{2021}).

\bibitem{bib2}
\bibinfo{author}{Alexander, D.~E.}
\newblock \bibinfo{title}{Social media in disaster risk reduction and crisis management}.
\newblock \emph{\bibinfo{journal}{Science and Engineering Ethics}} \textbf{\bibinfo{volume}{20}}, \bibinfo{pages}{717--733} (\bibinfo{year}{2014}).

\bibitem{bib3}
\bibinfo{author}{Berrios, R.}, \bibinfo{author}{Totterdell, P.} \& \bibinfo{author}{Kellett, S.}
\newblock \bibinfo{title}{Eliciting mixed emotions: a meta-analysis comparing models, types, and measures}.
\newblock \emph{\bibinfo{journal}{Frontiers in Psychology}} \textbf{\bibinfo{volume}{6}} (\bibinfo{year}{2015}).

\bibitem{bib41}
\bibinfo{author}{Siegel, D.~J.}
\newblock \emph{\bibinfo{title}{Mindsight: The new science of personal transformation}}  (\bibinfo{publisher}{Bantam}, \bibinfo{year}{2010}).

\bibitem{bib40}
\bibinfo{author}{Lund, R.}
\newblock \bibinfo{title}{Researching crisis—recognizing the unsettling experience of emotions}.
\newblock \emph{\bibinfo{journal}{Emotion, Space and Society}} \textbf{\bibinfo{volume}{5}}, \bibinfo{pages}{94--102} (\bibinfo{year}{2012}).

\bibitem{bib46}
\bibinfo{author}{Scherer, K.~R.}
\newblock \bibinfo{title}{What are emotions? and how can they be measured?}
\newblock \emph{\bibinfo{journal}{Social Science Information}} \textbf{\bibinfo{volume}{44}}, \bibinfo{pages}{695--729} (\bibinfo{year}{2005}).

\bibitem{bib42}
\bibinfo{author}{Ekman, P.}
\newblock \bibinfo{title}{An argument for basic emotions}.
\newblock \emph{\bibinfo{journal}{Cognition \& emotion}} \textbf{\bibinfo{volume}{6}}, \bibinfo{pages}{169--200} (\bibinfo{year}{1992}).

\bibitem{bib43}
\bibinfo{author}{Frijda, N.~H.}
\newblock \emph{\bibinfo{title}{The emotions}}  (\bibinfo{publisher}{Cambridge University Press}, \bibinfo{year}{1986}).

\bibitem{bib44}
\bibinfo{author}{Izard, C.~E.}
\newblock \bibinfo{title}{Basic emotions, relations among emotions, and emotion-cognition relations}  (\bibinfo{year}{1992}).

\bibitem{bib12}
\bibinfo{author}{Larsen, J.~T.} \& \bibinfo{author}{McGraw, A.~P.}
\newblock \bibinfo{title}{The case for mixed emotions}.
\newblock \emph{\bibinfo{journal}{Social and Personality Psychology Compass}} \textbf{\bibinfo{volume}{8}}, \bibinfo{pages}{263--274} (\bibinfo{year}{2014}).

\bibitem{bib10}
\bibinfo{author}{Moeller, J.}, \bibinfo{author}{Ivcevic, Z.}, \bibinfo{author}{Brackett, M.~A.} \& \bibinfo{author}{White, A.~E.}
\newblock \bibinfo{title}{Mixed emotions: Network analyses of intra-individual co-occurrences within and across situations}.
\newblock \emph{\bibinfo{journal}{Emotion}} \textbf{\bibinfo{volume}{18}}, \bibinfo{pages}{1106--1121} (\bibinfo{year}{2018}).

\bibitem{bib11}
\bibinfo{author}{De~Leersnyder, J.}, \bibinfo{author}{Koval, P.}, \bibinfo{author}{Kuppens, P.} \& \bibinfo{author}{Mesquita, B.}
\newblock \bibinfo{title}{Emotions and concerns: Situational evidence for their systematic co-occurrence.}
\newblock \emph{\bibinfo{journal}{Emotion}} \textbf{\bibinfo{volume}{18}}, \bibinfo{pages}{597--614} (\bibinfo{year}{2018}).

\bibitem{bib45}
\bibinfo{author}{Vansteelandt, K.}, \bibinfo{author}{Van~Mechelen, I.} \& \bibinfo{author}{Nezlek, J.~B.}
\newblock \bibinfo{title}{The co-occurrence of emotions in daily life: A multilevel approach}.
\newblock \emph{\bibinfo{journal}{Journal of Research in Personality}} \textbf{\bibinfo{volume}{39}}, \bibinfo{pages}{325--335} (\bibinfo{year}{2005}).

\bibitem{bib6}
\bibinfo{author}{Qi, X.}
\newblock \bibinfo{title}{Emotion in and through crisis}.
\newblock \emph{\bibinfo{journal}{The British Journal of Sociology}} \textbf{\bibinfo{volume}{75}}, \bibinfo{pages}{908--921} (\bibinfo{year}{2024}).

\bibitem{bib22}
\bibinfo{author}{Li, L.}, \bibinfo{author}{Wang, Z.}, \bibinfo{author}{Zhang, Q.} \& \bibinfo{author}{Wen, H.}
\newblock \bibinfo{title}{Effect of anger, anxiety, and sadness on the propagation scale of social media posts after natural disasters}.
\newblock \emph{\bibinfo{journal}{Information Processing \& Management}} \textbf{\bibinfo{volume}{57}}, \bibinfo{pages}{102313} (\bibinfo{year}{2020}).

\bibitem{bib47}
\bibinfo{author}{Kenett, Y.~N.}, \bibinfo{author}{Cardillo, E.~R.}, \bibinfo{author}{Christensen, A.~P.} \& \bibinfo{author}{Chatterjee, A.}
\newblock \bibinfo{title}{Aesthetic emotions are affected by context: a psychometric network analysis}.
\newblock \emph{\bibinfo{journal}{Scientific Reports}} \textbf{\bibinfo{volume}{13}} (\bibinfo{year}{2023}).

\bibitem{bib8}
\bibinfo{author}{Cooke, E.~M.}, \bibinfo{author}{Schuurman, N.~K.} \& \bibinfo{author}{Zheng, Y.}
\newblock \bibinfo{title}{Examining the within- and between-person structure of a short form of the positive and negative affect schedule: A multilevel and dynamic approach.}
\newblock \emph{\bibinfo{journal}{Psychological Assessment}} \textbf{\bibinfo{volume}{34}}, \bibinfo{pages}{1126--1137} (\bibinfo{year}{2022}).

\bibitem{bib9}
\bibinfo{author}{Lange, J.} \& \bibinfo{author}{Zickfeld, J.~H.}
\newblock \bibinfo{title}{Comparing implications of distinct emotion, network, and dimensional approaches for co-occurring emotions}.
\newblock \emph{\bibinfo{journal}{Emotion}} \textbf{\bibinfo{volume}{23}}, \bibinfo{pages}{2300--2321} (\bibinfo{year}{2023}).

\bibitem{bib68}
\bibinfo{author}{Huth, K. B.~S.}, \bibinfo{author}{Haslbeck, J. M.~B.}, \bibinfo{author}{Keetelaar, S.}, \bibinfo{author}{van Holst, R.~J.} \& \bibinfo{author}{Marsman, M.}
\newblock \bibinfo{title}{Statistical evidence in psychological networks}.
\newblock \emph{\bibinfo{journal}{Nature Human Behaviour}}  (\bibinfo{year}{2025}).

\bibitem{bib27}
\bibinfo{author}{Kuppens, P.} \& \bibinfo{author}{Verduyn, P.}
\newblock \bibinfo{title}{Emotion dynamics}.
\newblock \emph{\bibinfo{journal}{Current Opinion in Psychology}} \textbf{\bibinfo{volume}{17}}, \bibinfo{pages}{22--26} (\bibinfo{year}{2017}).

\bibitem{bib21}
\bibinfo{author}{Bringmann, L.~F.} \emph{et~al.}
\newblock \bibinfo{title}{Assessing temporal emotion dynamics using networks}.
\newblock \emph{\bibinfo{journal}{Assessment}} \textbf{\bibinfo{volume}{23}}, \bibinfo{pages}{425--435} (\bibinfo{year}{2016}).

\bibitem{bib60}
\bibinfo{author}{Fan, R.}, \bibinfo{author}{Zhao, J.}, \bibinfo{author}{Chen, Y.} \& \bibinfo{author}{Xu, K.}
\newblock \bibinfo{title}{Anger is more influential than joy: Sentiment correlation in weibo}.
\newblock \emph{\bibinfo{journal}{PLoS ONE}} \textbf{\bibinfo{volume}{9}}, \bibinfo{pages}{e110184} (\bibinfo{year}{2014}).

\bibitem{bib26}
\bibinfo{author}{Jin, Y.} \& \bibinfo{author}{Pang, A.}
\newblock \bibinfo{title}{Future directions of crisis communication research: Emotions in crisis–the next frontier}.
\newblock \emph{\bibinfo{journal}{The handbook of crisis communication}} \bibinfo{pages}{677--682} (\bibinfo{year}{2010}).

\bibitem{bib4}
\bibinfo{author}{Sano, Y.}, \bibinfo{author}{Takayasu, H.}, \bibinfo{author}{Havlin, S.} \& \bibinfo{author}{Takayasu, M.}
\newblock \bibinfo{title}{Identifying long-term periodic cycles and memories of collective emotion in online social media}.
\newblock \emph{\bibinfo{journal}{PLOS ONE}} \textbf{\bibinfo{volume}{14}}, \bibinfo{pages}{e0213843} (\bibinfo{year}{2019}).

\bibitem{bib5}
\bibinfo{author}{Wu, Q.}, \bibinfo{author}{Sano, Y.}, \bibinfo{author}{Takayasu, H.} \& \bibinfo{author}{Takayasu, M.}
\newblock \bibinfo{title}{Classification of endogenous and exogenous bursts in collective emotions based on weibo comments during covid-19}.
\newblock \emph{\bibinfo{journal}{Scientific Reports}} \textbf{\bibinfo{volume}{12}} (\bibinfo{year}{2022}).

\bibitem{bib7}
\bibinfo{author}{Brynielsson, J.}, \bibinfo{author}{Johansson, F.}, \bibinfo{author}{Jonsson, C.} \& \bibinfo{author}{Westling, A.}
\newblock \bibinfo{title}{Emotion classification of social media posts for estimating people’s reactions to communicated alert messages during crises}.
\newblock \emph{\bibinfo{journal}{Security Informatics}} \textbf{\bibinfo{volume}{3}} (\bibinfo{year}{2014}).

\bibitem{bib24}
\bibinfo{author}{An, L.}, \bibinfo{author}{An, N.}, \bibinfo{author}{Li, G.} \& \bibinfo{author}{Yu, C.}
\newblock \bibinfo{title}{Research on the dynamic mechanism of group emotional expression in the crisis}.
\newblock \emph{\bibinfo{journal}{Telematics and Informatics}} \textbf{\bibinfo{volume}{71}}, \bibinfo{pages}{101829} (\bibinfo{year}{2022}).

\bibitem{bib66}
\bibinfo{author}{Cao, G.} \emph{et~al.}
\newblock \bibinfo{title}{Analysis of social media data for public emotion on the wuhan lockdown event during the covid-19 pandemic}.
\newblock \emph{\bibinfo{journal}{Computer Methods and Programs in Biomedicine}} \textbf{\bibinfo{volume}{212}}, \bibinfo{pages}{106468} (\bibinfo{year}{2021}).

\bibitem{bib13}
\bibinfo{author}{Lorr, M.}, \bibinfo{author}{McNair, D.} \& \bibinfo{author}{Droppleman, L.}
\newblock \emph{\bibinfo{title}{Manual: Profile of Mood States. Educational and Industrial Testing Service}}  (\bibinfo{publisher}{CA Educational and Industrial Testing Service.}, \bibinfo{address}{San Diego}, \bibinfo{year}{1971}).

\bibitem{bib15}
\bibinfo{author}{Lochbaum, M.}, \bibinfo{author}{Zanatta, T.}, \bibinfo{author}{Kirschling, D.} \& \bibinfo{author}{May, E.}
\newblock \bibinfo{title}{The profile of moods states and athletic performance: A meta-analysis of published studies}.
\newblock \emph{\bibinfo{journal}{European Journal of Investigation in Health, Psychology and Education}} \textbf{\bibinfo{volume}{11}}, \bibinfo{pages}{50--70} (\bibinfo{year}{2021}).

\bibitem{bib16}
\bibinfo{author}{Cella, D.~F.} \emph{et~al.}
\newblock \bibinfo{title}{A brief poms measure of distress for cancer patients}.
\newblock \emph{\bibinfo{journal}{Journal of Chronic Diseases}} \textbf{\bibinfo{volume}{40}}, \bibinfo{pages}{939--942} (\bibinfo{year}{1987}).

\bibitem{bib48}
\bibinfo{author}{Baker, F.}, \bibinfo{author}{Denniston, M.}, \bibinfo{author}{Zabora, J.}, \bibinfo{author}{Polland, A.} \& \bibinfo{author}{Dudley, W.~N.}
\newblock \bibinfo{title}{A poms short form for cancer patients: psychometric and structural evaluation}.
\newblock \emph{\bibinfo{journal}{Psycho‐Oncology: Journal of the Psychological, Social and Behavioral Dimensions of Cancer}} \textbf{\bibinfo{volume}{11}}, \bibinfo{pages}{273--281} (\bibinfo{year}{2002}).

\bibitem{bib49}
\bibinfo{author}{Dilorenzo, T.~A.}, \bibinfo{author}{Bovbjerg, D.~H.}, \bibinfo{author}{Montgomery, G.~H.}, \bibinfo{author}{Valdimarsdottir, H.} \& \bibinfo{author}{Jacobsen, P.~B.}
\newblock \bibinfo{title}{The application of a shortened version of the profile of mood states in a sample of breast cancer chemotherapy patients}.
\newblock \emph{\bibinfo{journal}{British Journal of Health Psychology}} \textbf{\bibinfo{volume}{4}}, \bibinfo{pages}{315--325} (\bibinfo{year}{1999}).

\bibitem{bib50}
\bibinfo{author}{Renger, R.}
\newblock \bibinfo{title}{A review of the profile of mood states (poms) in the prediction of athletic success}.
\newblock \emph{\bibinfo{journal}{Journal of Applied Sport Psychology}} \textbf{\bibinfo{volume}{5}}, \bibinfo{pages}{78--84} (\bibinfo{year}{1993}).

\bibitem{bib51}
\bibinfo{author}{Terry, P.~C.}, \bibinfo{author}{Lane, A.~M.} \& \bibinfo{author}{Fogarty, G.~J.}
\newblock \bibinfo{title}{Construct validity of the profile of mood states—adolescents for use with adults}.
\newblock \emph{\bibinfo{journal}{Psychology of sport and exercise}} \textbf{\bibinfo{volume}{4}}, \bibinfo{pages}{125--139} (\bibinfo{year}{2003}).

\bibitem{bib36}
\bibinfo{author}{Plutchik, R.}
\newblock \emph{\bibinfo{title}{A general psychoevolutionary theory of emotion}}, \bibinfo{pages}{3--33} (\bibinfo{publisher}{Elsevier}, \bibinfo{year}{1980}).

\bibitem{bib58}
\bibinfo{author}{D’Mello, S.~K.} \& \bibinfo{author}{Graesser, A.~C.}
\newblock \emph{\bibinfo{title}{Confusion}}, \bibinfo{pages}{289--310} (\bibinfo{publisher}{Routledge}, \bibinfo{year}{2014}).

\bibitem{bib59}
\bibinfo{author}{St~Clair~Gibson, A.} \emph{et~al.}
\newblock \bibinfo{title}{The conscious perception of the sensation of fatigue}.
\newblock \emph{\bibinfo{journal}{Sports Medicine}} \textbf{\bibinfo{volume}{33}}, \bibinfo{pages}{167--176} (\bibinfo{year}{2003}).

\bibitem{bib52}
\bibinfo{author}{Kenett, Y.~N.}, \bibinfo{author}{Siew, C. S.~Q.} \& \bibinfo{author}{Vitevitch, M.~S.}
\newblock \bibinfo{title}{Network science in experimental psychology.}
\newblock \emph{\bibinfo{journal}{Canadian Journal of Experimental Psychology / Revue canadienne de psychologie expérimentale}} \textbf{\bibinfo{volume}{79}}, \bibinfo{pages}{1--3} (\bibinfo{year}{2025}).

\bibitem{bib53}
\bibinfo{author}{Siew, C. S.~Q.}, \bibinfo{author}{Wulff, D.~U.}, \bibinfo{author}{Beckage, N.~M.} \& \bibinfo{author}{Kenett, Y.~N.}
\newblock \bibinfo{title}{Cognitive network science: A review of research on cognition through the lens of network representations, processes, and dynamics}.
\newblock \emph{\bibinfo{journal}{Complexity}} \textbf{\bibinfo{volume}{2019}}, \bibinfo{pages}{1--24} (\bibinfo{year}{2019}).

\bibitem{bib28}
\bibinfo{author}{Bringmann, L.~F.} \emph{et~al.}
\newblock \bibinfo{title}{Assessing temporal emotion dynamics using networks}.
\newblock \emph{\bibinfo{journal}{Assessment}} \textbf{\bibinfo{volume}{23}}, \bibinfo{pages}{425--435} (\bibinfo{year}{2016}).

\bibitem{bib29}
\bibinfo{author}{Ryan, O.}, \bibinfo{author}{Dablander, F.} \& \bibinfo{author}{Haslbeck, J. M.~B.}
\newblock \bibinfo{title}{Toward a generative model for emotion dynamics.}
\newblock \emph{\bibinfo{journal}{Psychological Review}} \textbf{\bibinfo{volume}{132}}, \bibinfo{pages}{416--441} (\bibinfo{year}{2025}).

\bibitem{bib63}
\bibinfo{author}{Borsboom, D.} \emph{et~al.}
\newblock \bibinfo{title}{Network analysis of multivariate data in psychological science}.
\newblock \emph{\bibinfo{journal}{Nature Reviews Methods Primers}} \textbf{\bibinfo{volume}{1}} (\bibinfo{year}{2021}).

\bibitem{bib64}
\bibinfo{author}{Lunansky, G.} \emph{et~al.}
\newblock \bibinfo{title}{Bouncing back from life’s perturbations: Formalizing psychological resilience from a complex systems perspective.}
\newblock \emph{\bibinfo{journal}{Psychological Review}}  (\bibinfo{year}{2024}).

\bibitem{bib30}
\bibinfo{author}{Kenett, Y.~N.} \emph{et~al.}
\newblock \bibinfo{title}{Flexibility of thought in high creative individuals represented by percolation analysis}.
\newblock \emph{\bibinfo{journal}{Proceedings of the National Academy of Sciences}} \textbf{\bibinfo{volume}{115}}, \bibinfo{pages}{867--872} (\bibinfo{year}{2018}).

\bibitem{bib32}
\bibinfo{author}{Kenett, Y.~N.}, \bibinfo{author}{Anaki, D.} \& \bibinfo{author}{Faust, M.}
\newblock \bibinfo{title}{Investigating the structure of semantic networks in low and high creative persons}.
\newblock \emph{\bibinfo{journal}{Frontiers in human neuroscience}} \textbf{\bibinfo{volume}{8}}, \bibinfo{pages}{407} (\bibinfo{year}{2014}).

\bibitem{bib54}
\bibinfo{author}{Hills, T.~T.} \& \bibinfo{author}{Kenett, Y.~N.}
\newblock \bibinfo{title}{Is the mind a network? maps, vehicles, and skyhooks in cognitive network science}.
\newblock \emph{\bibinfo{journal}{Topics in Cognitive Science}} \textbf{\bibinfo{volume}{14}}, \bibinfo{pages}{189--208} (\bibinfo{year}{2022}).

\bibitem{bib55}
\bibinfo{author}{Kenett, Y.~N.}
\newblock \bibinfo{title}{The role of knowledge in creative thinking}.
\newblock \emph{\bibinfo{journal}{Creativity Research Journal}} \textbf{\bibinfo{volume}{37}}, \bibinfo{pages}{242--249} (\bibinfo{year}{2025}).

\bibitem{bib33}
\bibinfo{author}{Beck, E.~D.} \& \bibinfo{author}{Jackson, J.~J.}
\newblock \emph{\bibinfo{title}{Network approaches to representing and understanding personality dynamics}}, \bibinfo{pages}{465--497} (\bibinfo{publisher}{Elsevier}, \bibinfo{year}{2021}).

\bibitem{bib34}
\bibinfo{author}{Jayawickreme, E.}, \bibinfo{author}{Fleeson, W.}, \bibinfo{author}{Beck, E.~D.}, \bibinfo{author}{Baumert, A.} \& \bibinfo{author}{Adler, J.~M.}
\newblock \bibinfo{title}{Personality dynamics}.
\newblock \emph{\bibinfo{journal}{Personality Science}} \textbf{\bibinfo{volume}{2}}, \bibinfo{pages}{e6179} (\bibinfo{year}{2021}).

\bibitem{bib56}
\bibinfo{author}{Chen, Q.} \emph{et~al.}
\newblock \bibinfo{title}{Mapping the creative personality: A psychometric network analysis of highly creative artists and scientists}.
\newblock \emph{\bibinfo{journal}{Creativity Research Journal}} \textbf{\bibinfo{volume}{35}}, \bibinfo{pages}{455--470} (\bibinfo{year}{2023}).
\newblock \bibinfo{note}{Doi: 10.1080/10400419.2023.2184558}.

\bibitem{bib62}
\bibinfo{author}{Yee, J.~L.} \& \bibinfo{author}{Niemeier, D.}
\newblock \bibinfo{title}{Advantages and disadvantages: Longitudinal vs. repeated cross-section surveys}  (\bibinfo{year}{1996}).

\bibitem{bib14}
\bibinfo{author}{Levin, K.~A.}
\newblock \bibinfo{title}{Study design iii: Cross-sectional studies}.
\newblock \emph{\bibinfo{journal}{Evidence-Based Dentistry}} \textbf{\bibinfo{volume}{7}}, \bibinfo{pages}{24--25} (\bibinfo{year}{2006}).

\bibitem{bib57}
\bibinfo{author}{Benjamini, Y.} \& \bibinfo{author}{Hochberg, Y.}
\newblock \bibinfo{title}{Controlling the false discovery rate: a practical and powerful approach to multiple testing}.
\newblock \emph{\bibinfo{journal}{Journal of the Royal statistical society: series B (Methodological)}} \textbf{\bibinfo{volume}{57}}, \bibinfo{pages}{289--300} (\bibinfo{year}{1995}).

\bibitem{bib65}
\bibinfo{author}{Storey, J.~D.}
\newblock \emph{\bibinfo{title}{False discovery rate}}, \bibinfo{pages}{504--508} (\bibinfo{publisher}{Springer}, \bibinfo{year}{2011}).

\bibitem{bib38}
\bibinfo{author}{Di~Martino, E.}, \bibinfo{author}{Galeazzi, A.}, \bibinfo{author}{Starnini, M.}, \bibinfo{author}{Quattrociocchi, W.} \& \bibinfo{author}{Cinelli, M.}
\newblock \bibinfo{title}{Ideological fragmentation of the social media ecosystem: From echo chambers to echo platforms}.
\newblock \emph{\bibinfo{journal}{arXiv preprint arXiv:2411.16826}}  (\bibinfo{year}{2024}).

\bibitem{bib39}
\bibinfo{author}{Salloum, A.}, \bibinfo{author}{Quelle, D.}, \bibinfo{author}{Iannucci, L.}, \bibinfo{author}{Bovet, A.} \& \bibinfo{author}{Kivelä, M.}
\newblock \bibinfo{title}{Politics and polarization on bluesky}.
\newblock \emph{\bibinfo{journal}{arXiv preprint arXiv:2506.03443}}  (\bibinfo{year}{2025}).

\bibitem{bib37}
\bibinfo{author}{Watson, D.}, \bibinfo{author}{Clark, L.~A.} \& \bibinfo{author}{Tellegen, A.}
\newblock \bibinfo{title}{Development and validation of brief measures of positive and negative affect: the panas scales}.
\newblock \emph{\bibinfo{journal}{Journal of personality and social psychology}} \textbf{\bibinfo{volume}{54}}, \bibinfo{pages}{1063} (\bibinfo{year}{1988}).

\bibitem{bib35}
\bibinfo{author}{Lovibond, P.~F.} \& \bibinfo{author}{Lovibond, S.~H.}
\newblock \bibinfo{title}{Depression anxiety and stress scales}.
\newblock \emph{\bibinfo{journal}{Behaviour Research and Therapy}}  (\bibinfo{year}{1995}).

\bibitem{bib17}
\bibinfo{author}{Tamura, K.}, \bibinfo{author}{Sano, Y.} \& \bibinfo{author}{Shiozaki, J.}
\newblock \bibinfo{title}{Quantifying collective emotions: Japan’s societal trends through enhanced sentiment index using poms2 and sns}.
\newblock \emph{\bibinfo{journal}{2024 IEEE International Conference on Big Data (BigData)}} \bibinfo{pages}{3082--3087} (\bibinfo{year}{2024}).

\bibitem{bib19}
\bibinfo{author}{Dover, Y.} \& \bibinfo{author}{Moore, Z.}
\newblock \bibinfo{title}{Using free association networks to extract characteristic patterns of affect dynamics}.
\newblock \emph{\bibinfo{journal}{Proceedings of the Royal Society A: Mathematical, Physical and Engineering Sciences}} \textbf{\bibinfo{volume}{476}}, \bibinfo{pages}{20190647} (\bibinfo{year}{2020}).

\bibitem{bib20}
\bibinfo{author}{Li, Y.}, \bibinfo{author}{Masitah, A.} \& \bibinfo{author}{Hills, T.~T.}
\newblock \bibinfo{title}{The emotional recall task: Juxtaposing recall and recognition-based affect scales.}
\newblock \emph{\bibinfo{journal}{Journal of Experimental Psychology: Learning, Memory, and Cognition}} \textbf{\bibinfo{volume}{46}}, \bibinfo{pages}{1782--1794} (\bibinfo{year}{2020}).

\bibitem{bib18}
\bibinfo{author}{Matsuno, S.}, \bibinfo{author}{Mizuki, S.} \& \bibinfo{author}{Sakaki, T.}
\newblock \bibinfo{title}{Constructing of the word embedding model by japanese large scale sns + web corpus}.
\newblock \emph{\bibinfo{journal}{Proceedings of the Annual Conference of JSAI}} \textbf{\bibinfo{volume}{JSAI2019}}, \bibinfo{pages}{4Rin113--4Rin113} (\bibinfo{year}{2019}).

\bibitem{bib67}
\bibinfo{author}{Kenett, Y.~N.} \emph{et~al.}
\newblock \bibinfo{title}{Flexibility of thought in high creative individuals represented by percolation analysis}.
\newblock \emph{\bibinfo{journal}{Proceedings of the National Academy of Sciences}} \textbf{\bibinfo{volume}{115}}, \bibinfo{pages}{867--872} (\bibinfo{year}{2018}).

\bibitem{bib61}
\bibinfo{author}{Levy, O.} \emph{et~al.}
\newblock \bibinfo{title}{Unveiling the nature of interaction between semantics and phonology in lexical access based on multilayer networks}.
\newblock \emph{\bibinfo{journal}{Scientific Reports}} \textbf{\bibinfo{volume}{11}} (\bibinfo{year}{2021}).

\end{thebibliography}

\newpage

\appendix


\section*{Supplementary material}

# Supplementary Information: Stable Emotional Co-occurrence Patterns Revealed by Network Analysis of Social Media

Qianyun Wu, Orr Levy, Yoed Kenett, Yukie Sano, Hideki Takayasu, Shlomo Havlin, Misako Takayasu

## 1 Rank of emotion links across time within datasets

In Fig. S1 we show the heatmap of emotion links over time within each period. Visually, we can see that the rank of emotion links exhibits high stability across time. Interestingly, we can observe a change of rank before and after the earthquake taking place on 11th March 2011, also before and after the start of vaccination in April 2021.

To further quantitatively measure the stability of emotion link's ranks across time, we show **Fig. 2** and Fig. S2- The Spearman correlation of emotion networks across time within each dataset.

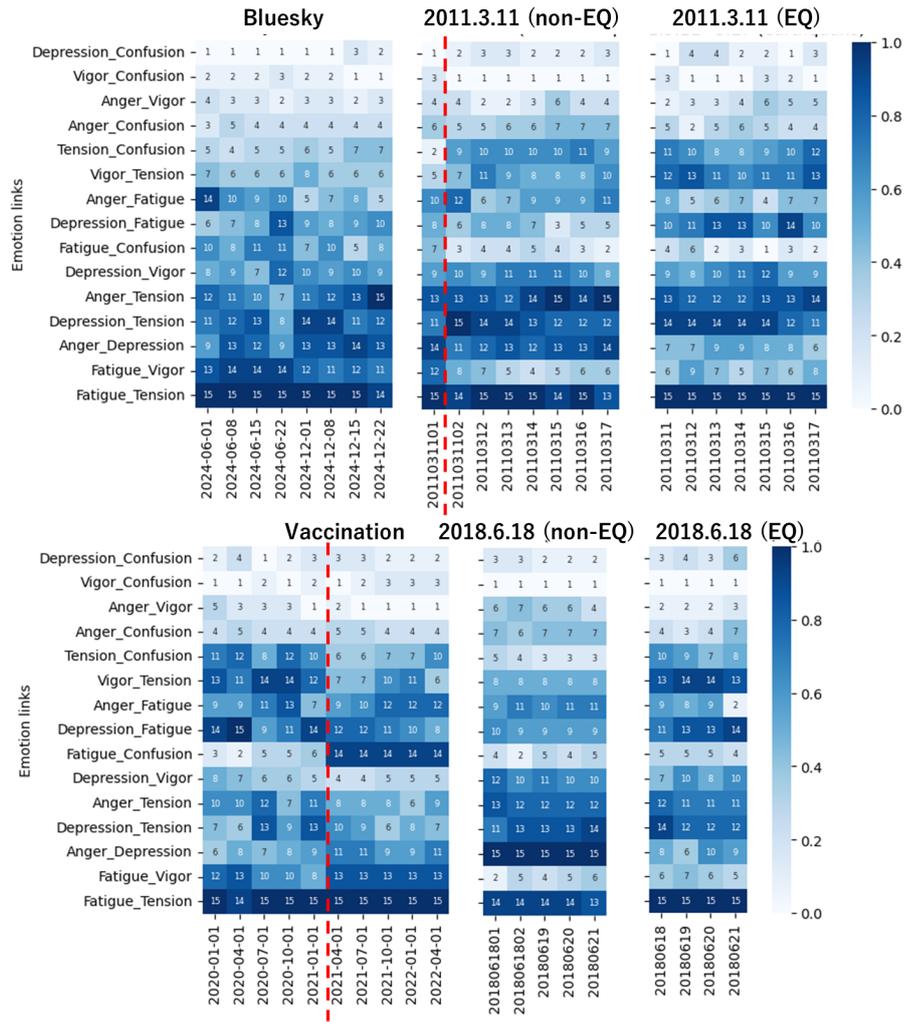

Supplementary Fig. S1: Heatmap showing the ranks of emotion links over time within each dataset. The numerical annotations indicate the ranks of emotion link strengths in descending order, where rank 1 represents the strongest link.

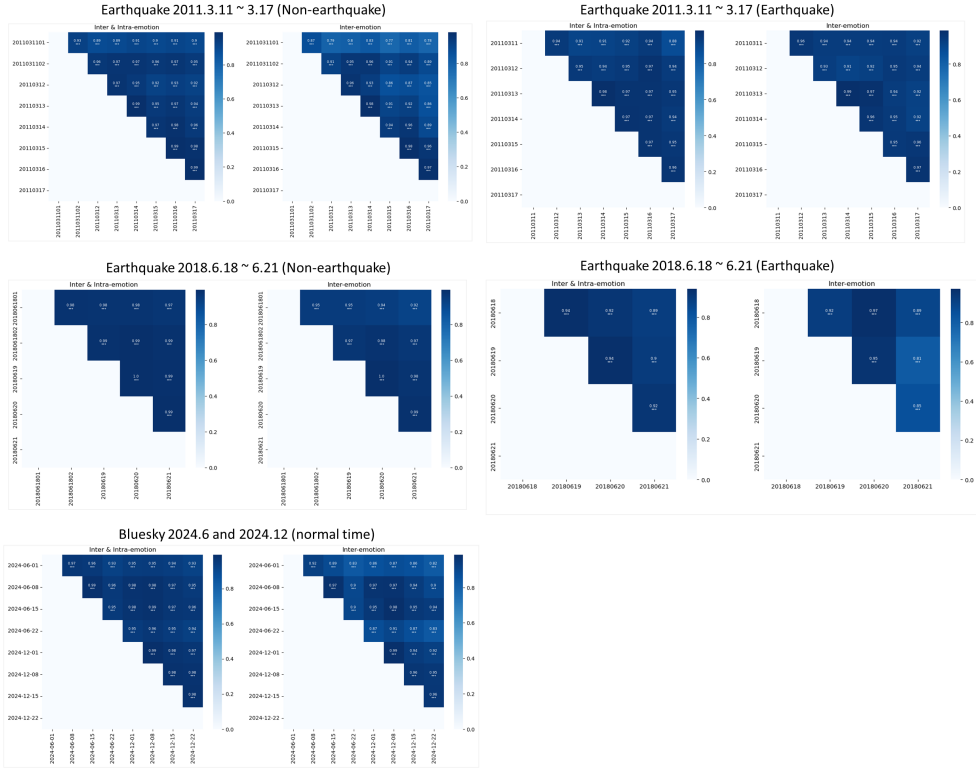

**Supplementary Fig. S2**: Heatmap of Spearman rank correlation of emotion link's strength between emotion network snapshots within each dataset (earthquake 2011.3.11, earthquake 2018.6.18, Bluesky 2024.6 and 2024.12). Each sub-figure includes a heatmap comparing both intra- and inter-emotion links (left) and a heatmap comparing inter-emotion links (right). For all datasets, we adopt thresholds of link weight $\geq$ top 10% and link strength $\geq$ 3 (see *Methods*). The color bar to the right of each subplot shows the color density of the corresponding value, ranging from 0 to 1. The annotations indicate Spearman's rank correlation coefficient, along with the corresponding $p$-values and significance asterisks.

# 2 Jaccard Index of temporal emotion concept (word) networks within and across datasets.

To compare the similarity between concept networks constructed from significant concept links, we adopt the Jaccard Index. For two sets of significant concept links $X_{t_1}$ and $X_{t_2}$ in two time windows $t_1$ and $t_2$, the Jaccard Index represents the fraction of number of elements in the intersection $X_{t1} \cap X_{t2}$ over the number of elements in the union $X_{t1} \cup X_{t2}$ (Equation S1). The Jaccard Index ranges from 0 to 1, where 0 indicates no shared concept links, and 1 indicates complete overlap between the two concept networks.

Fig. S3 presents the heatmap of the Jaccard Index measuring the similarity of concept networks across different time windows for each dataset. In Table. S1, we show the mean Jaccard Index ± the corresponding standard deviation. The similarity values are relatively low as compared to the high similarity between emotion networks, indicating a relatively high variation in the co-occurrence of emotion concepts over time. Based on the density of color patterns, the heatmap visually clusters into two distinct periods—before and after the events, such as the start of COVID-19 vaccination in April 2021 and the occurrence of the earthquake. This shift suggests that Twitter users exhibit dynamic patterns in emotional expression in response to external events such as the rollout of vaccination. We can also observe a higher correlation between adjacent time windows, suggesting that the usage of emotion concepts have a memory effect.

$$J(X_{t_1}, X_{t_2}) = \frac{|X_{t_1} \cap X_{t_2}|}{|X_{t_1} \cup X_{t_2}|} \tag{1}$$

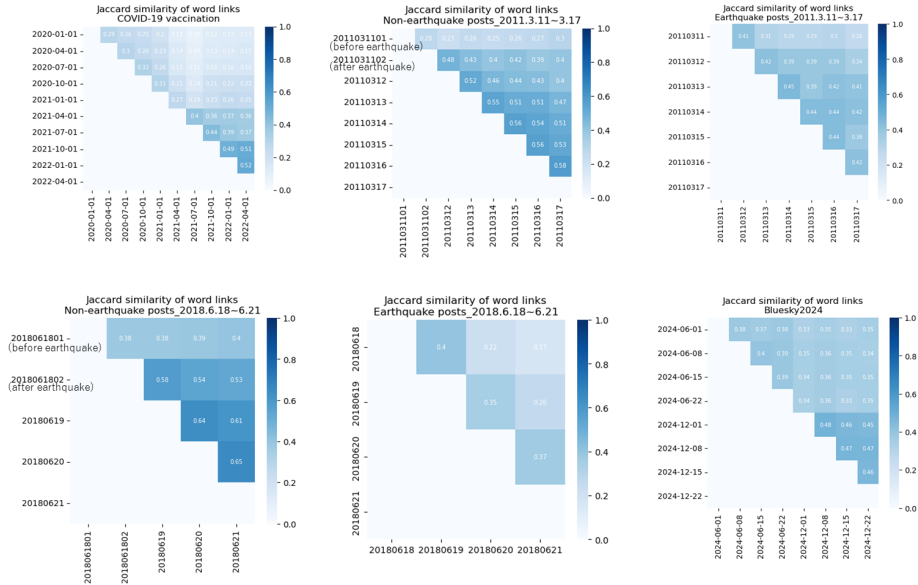

**Supplementary Fig. S3**: Heatmap of Jaccard Index between temporal concept (word) network snapshots within each dataset (earthquake 2011.3.11, earthquake 2018.6.18, Bluesky 2024.6 and 2024.12). The color bar to the right of each subplot shows the color density of the corresponding value, ranging from 0 to 1. The annotations indicate Jaccard Index comparing the similarity of significant concept links between two time windows.

**Supplementary Table. S1**: Summary of Jaccard similarity within crisis datasets

|  | Pre-vaccination | Vaccination | All tweets (2011.3.11-3.17) | Earthquake tweets (2011.3.11-3.17) | All tweets (2018.6.18-6.21) | Earthquake tweets (2018.6.18-6.21) |
|---|---|---|---|---|---|---|
| Jaccard Index | 0.27 ±0.04 | 0.42 ±0.06 | 0.42 ±0.12 | 0.38 ±0.06 | 0.49 ±0.11 | 0.30 ±0.08 |

We further assess the stability of concept links from a different perspective—the frequency with which a link appears as significant across time windows within each dataset (Fig. S4). Taking the vaccination dataset as an example, some concept links may be significant in 5 out of 10 quarters, while others appear only once. To evaluate whether such recurrence exceeds what would be expected by chance, we compare the observed frequency of each concept link's significance with a null distribution generated via random sampling.

Specifically, we perform 1,000 random samplings. In each sampling, we randomly select, from the full set of possible concept links, the same number of links as observed to be significant in each time window. We then calculate how often each link appears across all time windows. For each recurrence level (e.g., 5 times), we compute the cumulative distribution function (CDF) representing the probability that a link recurs in at least $r$ time windows: $F'(\geq r)$ for the actual data and $F(\geq r)$ for the random data. We then calculate the $p$-value as $p = \frac{F(\geq r)}{F'(\geq r)}$, which indicates the proportion of recurrence in the random case relative to the actual case. A $p$-value below 0.05 indicates that a concept link is significantly more stable than would be expected by chance.

In Fig. S4, we show the comparison between actual significant recurrence frequencies and the random baseline for each dataset. In each sub-figure, the sub-figure on the left shows the frequency of concept links with different recurrence levels under random (blue) and actual (red) cases, while the sub-figure on the right shows the CDF of significance. The black annotations indicate the $p$-value at different recurrence levels, while the red annotations indicate the proportion of concept links that are significantly more stable than random ($p < 0.05$).

Across datasets, we observe that 21% to 54% of concept links exhibit statistically significant stability over time, suggesting that only a modest fraction of word associations consistently recur throughout the study period.

Similarly, we test the significance of concept link stability across datasets by comparing the recurrence of each concept link across all datasets with that in a randomized baseline. Specifically, we first compute the number of concept links that appear significant in varying numbers of datasets. To generate the baseline, we perform 1,000 random re-samplings, where in each sample, we randomly select from all possible concept links the same number of significant links as observed in each dataset. This yields the distribution of recurrence levels under the null model. In Fig. S5, we present both the histogram and CDF of concept link recurrence across datasets for actual and random data. We observe that only a relative small fraction of word links (39%) appear significantly more stable than random.

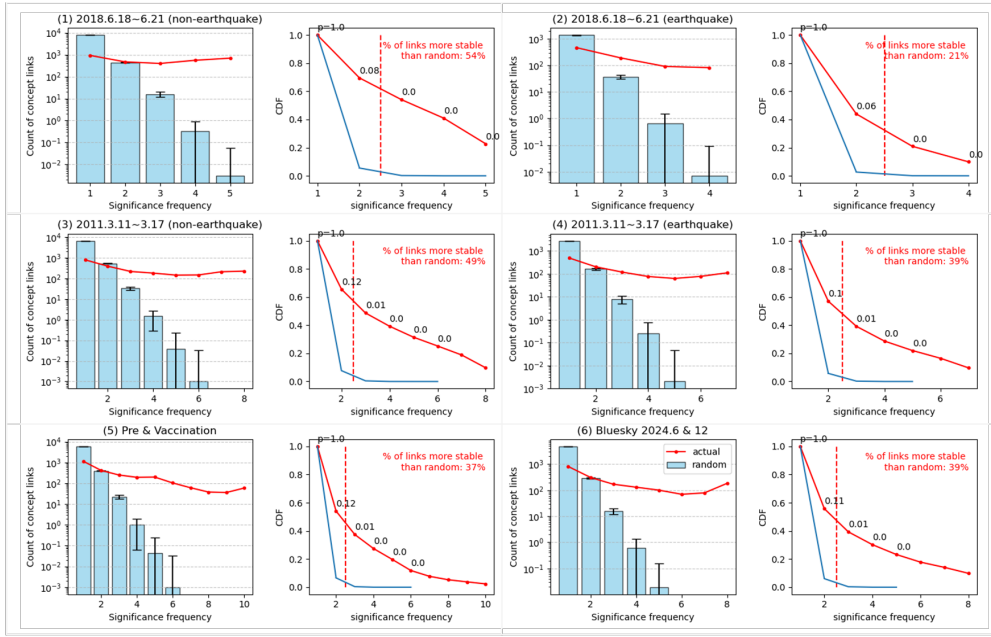

**Supplementary Fig. S4**: Histogram and cumulative distribution of significance frequency of concept links. The figure shows how often each concept link is identified as significant across time windows within a dataset. In each sub-figure, the left sub-figure depicts the histogram of frequency while the right sub-figure shows the corresponding CDF. The x-axis represents the number of time windows in which a concept link is significant, and the y-axis indicates the number (left sub-figure) or the cumulative distribution of concept links at each frequency level. The red line shows the actual distribution of significant concept links. The blue bar chart or line chart with error bars represents the mean ± standard deviation from 1,000 random re-samplings, where in each sampling, a number (same to the number of significant links) of concept links are randomly selected from the full set of possible concept links in each time window, and their recurrence frequencies are calculated. The black annotations indicate the $p$-value, and the red annotations indicate the proportion of concept links that are significantly more stable than random.

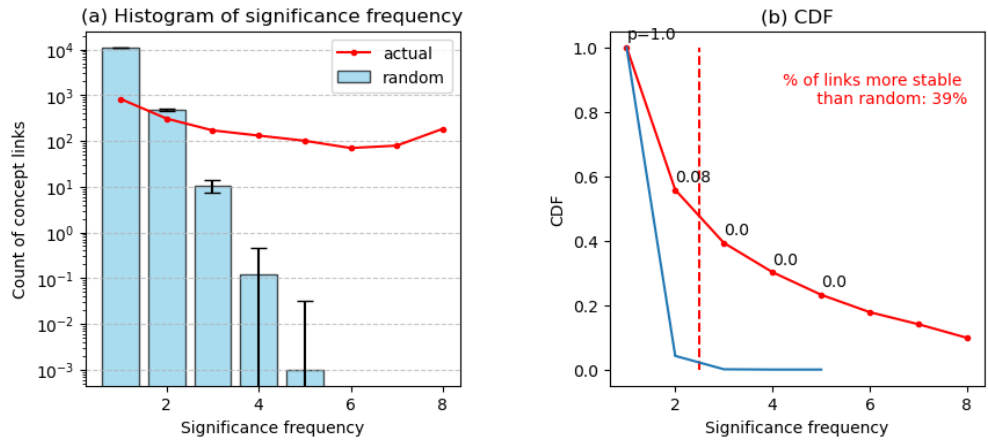

**Supplementary Fig. S5**: Histogram and cumulative distribution of significance frequency of concept links. (a) Histogram of significance frequency. The x-axis represents the number of time windows in which a concept link is significant, and the y-axis indicates the number of concept links at each frequency level. (b) CDF of significance frequency. The black annotations indicate the $p$-value, and the red annotations indicate the proportion of concept links that are significantly more stable than random.

# 3 Stability of emotion networks over time within each dataset based on the strength of emotion links.

In the Results section, we showed the stability of emotion networks across time within each dataset using the Spearman correlation, which evaluates the rank of emotion links' strengths. Here, instead of comparing the ranks, we further compare the strength of emotion links to show that they also exhibit a high degree of similarity over time.

In Fig S6, we show a strong linear relationship between the strength of emotion links in successive time windows, which indicates that the structure of emotion co-occurrence is highly stable over time, with links maintaining similar relative strengths across periods.

In the Results section, we employ Spearman's rank correlation to measure the similarity between emotion networks instead of Pearson's correlation. This choice is made because of the distribution of emotion link's strengths, as illustrated in **Fig. 4**, where certain emotion links—such as Fatigue–Fatigue—are disproportionately dominant compared to others. Such skewness could introduce bias when using Pearson's correlation, which is sensitive to extreme values. For completeness, in Fig. S7 we present the results based on Pearson's correlation, which similarly demonstrate a high level of similarity across temporal emotion networks.

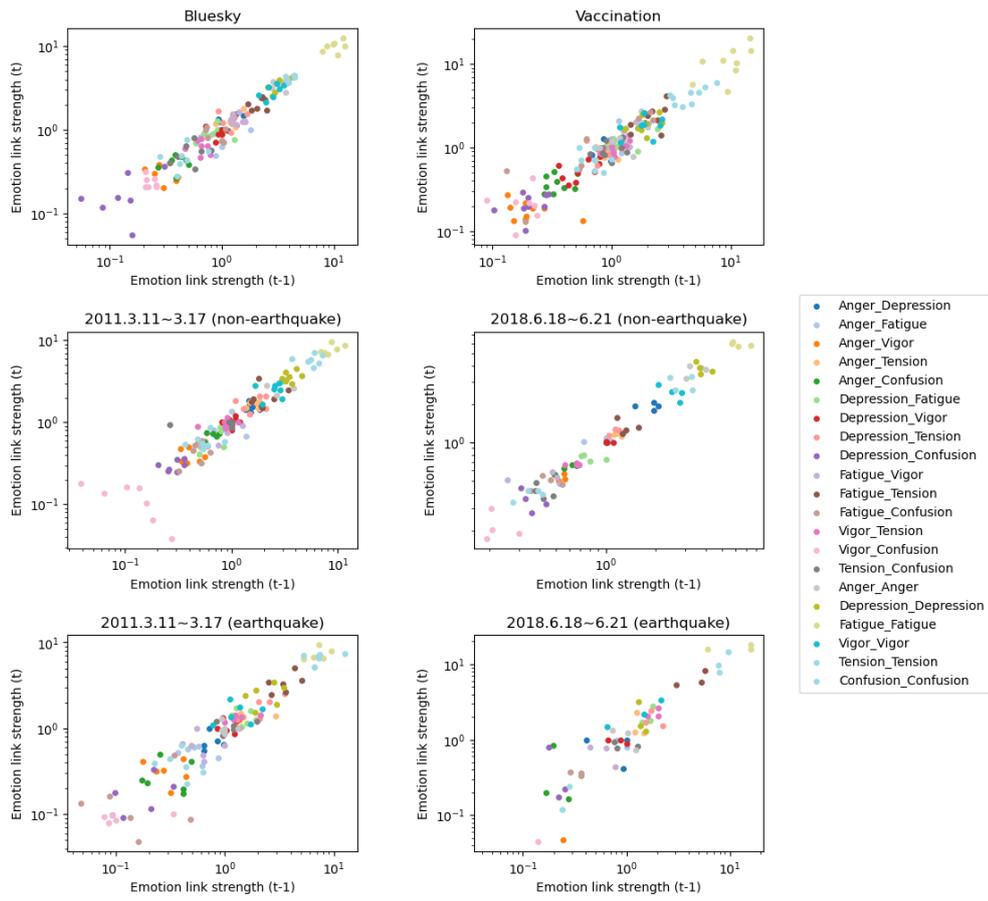

**Supplementary Fig. S6**: Scatter plots comparing the strength of emotion link in the previous time stamp $(t-1)$ to the next time stamp $(t)$. Sub-figures show the result for the respective datasets. Each dot represents a pair of strengths of the same emotion link in adjacent time windows.

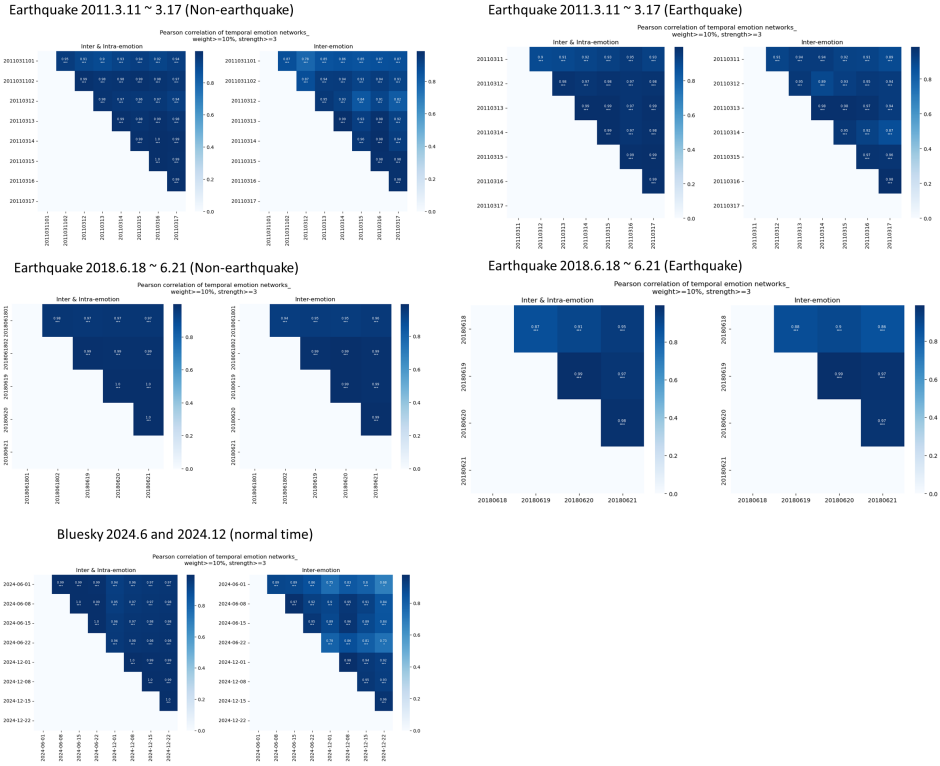

**Supplementary Fig. S7**: Heatmap of Pearson's correlation of emotion link's strength between emotion network snapshots within each dataset (earthquake 2011.3.11, earthquake 2018.6.18, Bluesky 2024.6 and 2024.12). Each sub-figure includes a heatmap comparing both intra- and inter-emotion links (left) and a heatmap comparing inter-emotion links (right). For all datasets, we adopt thresholds of link weight top 10% and link strength 3 (see Methods). The color bar to the right of each subplot shows the color density of the corresponding value, ranging from 0 to 1. The annotations indicate Pearson's correlation coefficient, along with the corresponding p-values and significance asterisks.

# 4 The average and standard deviation of Spearman's correlation comparing emotion networks across dataset

Table S2 shows the average and standard deviation of Spearman's correlation between emotion networks across datasets. The values above diagonals represent shows Spearman correlation for both the inter- & intra-emotion links, while the values below diagonals shows the correlation for only inter-emotion links.

**Supplementary Table. S2**: The average and standard deviation of Spearman's correlation comparing emotion networks across dataset.

| Datasets | Bluesky | 2011.3 (non-EQ) | 2018.6 (non-EQ) | 2011.3 (EQ) | 2018.6 (EQ) | pre-vaccine | vaccine |
|---|---|---|---|---|---|---|---|
| Bluesky |  | 0.87 ±0.06 | 0.86 ±0.04 | 0.75 ±0.07 | 0.73 ±0.04 | 0.79 ±0.05 | 0.84 ±0.05 |
| 2011.3 (non-EQ) | 0.72 ±0.14 |  | 0.91 ±0.02 | 0.81 ±0.08 | 0.76 ±0.06 | 0.79 ±0.06 | 0.68 ±0.07 |
| 2018.6 (non-EQ) | 0.69 ±0.11 | 0.83 ±0.03 |  | 0.75 ±0.07 | 0.74 ±0.04 | 0.73 ±0.05 | 0.65 ±0.07 |
| 2011.3 (EQ) | 0.64 ±0.11 | 0.76 ±0.1 | 0.69 ±0.08 |  | 0.9 ±0.03 | 0.84 ±0.05 | 0.59 ±0.07 |
| 2018.6 (EQ) | 0.61 ±0.11 | 0.7 ±0.12 | 0.67 ±0.06 | 0.89 ±0.05 |  | 0.83 ±0.05 | 0.64 ±0.06 |
| pre-vaccine | 0.62 ±0.09 | 0.62 ±0.11 | 0.52 ±0.09 | 0.79 ±0.08 | 0.8 ±0.11 |  | 0.74 ±0.04 |
| vaccine | 0.73 ±0.09 | 0.44 ±0.11 | 0.39 ±0.08 | 0.41 ±0.1 | 0.5 ±0.12 | 0.62 ±0.06 |  |

# 5 Temporal rescaled strength of emotion links

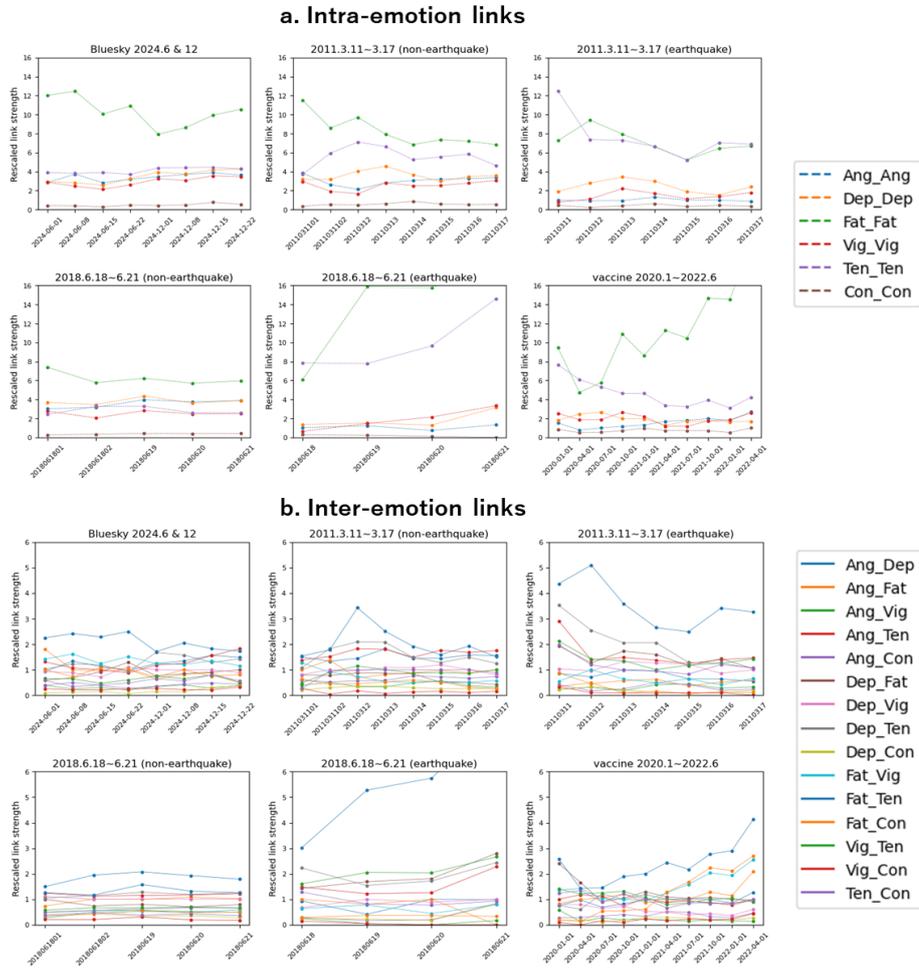

**Supplementary Fig. S8**: The temporal strengths (rescaled) of emotion links for each dataset. (a) The strength of intra-emotion links over time. (b) The strength of inter-emotion links over time.

# 6 Statistics of the two-sample t-test for comparing emotion link strength across datasets

To investigate how do the strengths of emotion links change across different situations and if these changes are significant, we run a two-sample t-test to compare two sets of link strengths from different datasets $\{S'^{(t)}_{ij}\}_{\text{dataset1}}$ and $\{S'^{(t)}_{ij}\}_{\text{dataset2}}$, where $i$ and $j$ represent emotions, and $t$ represents time window of a dataset. Table A3. reports the statistics of the two-sample t-test. where $t$(degree of freedom) represents the $t$-statistics, $p$ represents p-value and $d$ represents Cohen's $d$.

**Supplementary Table. S3**: Two-sample t-test for each emotion link between datasets.

| Emotion Links | EQ2011(non-EQ) vs. BS | EQ2018(non-EQ) vs. BS | EQ2011(EQ) vs. BS | EQ2018(EQ) vs. BS | Pre-Vaccine vs. BS | Vaccine vs. BS | EQ2018(non-EQ) vs. EQ2018(EQ) | EQ2011(non-EQ) vs. EQ2011(EQ) | Vaccine vs. Pre-Vaccine |
|---|---|---|---|---|---|---|---|---|---|
| Anger-Anger | t(14) = -1.59 p=0.1339 d=-0.85 | t(11) = 0.64 p=0.5344 d=0.4 | t(13) = -14.82 p=0.0 d=-8.21 | t(10) = -10.26 p=0.0 d=-6.79 | t(11) = -10.76 p=0.0001 d=-6.62 | t(11) = -6.04 p=0.0001 d=-3.74 | t(7) = -9.95 p=0.0 d=-7.52 | t(13) = -10.05 p=0.0 d=-5.56 | t(8) = 3.74 p=0.0057 d=2.64 |
| Anger-Depression | t(14) = 2.79 p=0.0143 d=1.49 | t(11) = 4.64 p=0.0007 d=2.87 | t(13) = -4.55 p=0.0005 d=-2.53 | t(10) = -2.77 p=0.0197 d=-1.87 | t(11) = -3.62 p=0.004 d=-2.22 | t(11) = -1.92 p=0.0809 d=-1.18 | t(7) = -6.07 p=0.0005 d=-4.63 | t(13) = -9.11 p=0.0 d=-5.07 | t(8) = 2.09 p=0.0697 d=1.48 |
| Anger-Fatigue | t(14) = 0.01 p=0.9895 d=0.01 | t(11) = 0.12 p=0.9084 d=0.07 | t(13) = -2.37 p=0.0339 d=-1.31 | t(10) = -0.96 p=0.3581 d=-0.65 | t(11) = 0.23 p=0.8211 d=0.14 | t(11) = -0.61 p=0.1898 d=0.87 | t(7) = -1.22 p=0.2635 d=-0.94 | t(13) = -3.92 p=0.0018 d=-2.18 | t(8) = 1.32 p=0.2242 d=0.93 |
| Anger-Vigor | t(14) = 2.26 p=0.0402 d=1.21 | t(11) = 5.72 p=0.0001 d=3.5 | t(13) = -0.17 p=0.8689 d=-0.09 | t(10) = -3.81 p=0.0034 d=-2.59 | t(11) = -0.61 p=0.5574 d=-0.38 | t(11) = -4.48 p=0.0009 d=-2.74 | t(7) = -7.79 p=0.0001 d=-6.01 | t(13) = -2.13 p=0.0532 d=-1.18 | t(8) = -1.46 p=0.1834 d=-1.03 |
| Anger-Tension | t(14) = 3.19 p=0.0066 d=1.7 | t(11) = -0.54 p=0.597 d=-0.33 | t(13) = 1.49 p=0.1596 d=0.83 | t(10) = 1.32 p=0.2168 d=0.9 | t(11) = -1.61 p=0.1368 d=-0.98 | t(11) = -2.6 p=0.0246 d=-1.59 | t(7) = 1.73 p=0.1268 d=1.34 | t(13) = -0.21 p=0.8358 d=-0.12 | t(8) = -1.49 p=0.1734 d=-1.06 |
| Anger-Confusion | t(14) = 5.72 p=0.0001 d=3.06 | t(11) = 4.62 p=0.0007 d=2.85 | t(13) = -1.8 p=0.095 d=-1.0 | t(10) = -0.33 p=0.7471 d=-0.23 | t(11) = -2.09 p=0.0601 d=-1.29 | t(11) = 0.18 p=0.8587 d=0.11 | t(7) = -1.62 p=0.1486 d=-1.25 | t(13) = -5.92 p=0.0001 d=-3.3 | t(8) = 1.7 p=0.1272 d=1.2 |
| Depression-Depression | t(14) = 0.4 p=0.6967 d=0.21 | t(11) = 1.05 p=0.3171 d=0.64 | t(13) = -3.03 p=0.0097 d=-1.68 | t(10) = -3.65 p=0.0045 d=-2.46 | t(11) = -3.99 p=0.0021 d=-2.45 | t(11) = -6.07 p=0.0001 d=-3.7 | t(7) = -4.65 p=0.0023 d=-3.58 | t(13) = -3.73 p=0.0025 d=-2.08 | t(8) = -2.98 p=0.0175 d=-2.11 |
| Depression-Fatigue | t(14) = -1.75 p=0.1022 d=-0.93 | t(11) = -0.63 p=0.5399 d=-0.39 | t(13) = 4.11 p=0.0012 d=2.29 | t(10) = 4.59 p=0.001 d=3.19 | t(11) = 2.56 p=0.0264 d=1.62 | t(11) = 1.61 p=0.1359 d=0.98 | t(7) = 4.15 p=0.0043 d=3.21 | t(13) = 5.6 p=0.0001 d=3.12 | t(8) = -1.58 p=0.1529 d=-1.12 |
| Depression-Vigor | t(14) = 0.77 p=0.4517 d=0.41 | t(11) = 1.57 p=0.1457 d=0.96 | t(13) = 2.05 p=0.061 d=1.14 | t(10) = -0.65 p=0.5327 d=-0.44 | t(11) = -3.7 p=0.0 d=-2.29 | t(11) = -7.66 p=0.0 d=-4.73 | t(7) = -1.89 p=0.1011 d=-1.46 | t(13) = 1.32 p=0.2097 d=0.73 | t(8) = -3.42 p=0.0091 d=-2.42 |
| Depression-Tension | t(14) = 1.72 p=0.1076 d=0.92 | t(11) = -0.9 p=0.3876 d=-0.55 | t(13) = 2.1 p=0.0557 d=1.17 | t(10) = 3.55 p=0.0053 d=2.42 | t(11) = -2.29 p=0.0428 d=-1.42 | t(11) = -3.1 p=0.0101 d=-1.89 | t(7) = 4.18 p=0.0041 d=3.23 | t(13) = 1.17 p=0.2647 d=0.65 | t(8) = -0.28 p=0.7835 d=-0.2 |
| Depression-Confusion | t(14) = 2.67 p=0.0183 d=1.43 | t(11) = 3.14 p=0.0093 d=1.93 | t(13) = 0.11 p=0.9177 d=0.06 | t(10) = 1.69 p=0.1225 d=1.17 | t(11) = 0.3 p=0.7731 d=0.18 | t(11) = 1.34 p=0.206 d=0.82 | t(7) = 0.17 p=0.8735 d=0.13 | t(13) = -2.98 p=0.0107 d=-1.66 | t(8) = 1.47 p=0.1788 d=1.04 |
| Fatigue-Fatigue | t(14) = -2.59 p=0.0212 d=-1.39 | t(11) = -5.52 p=0.0002 d=-3.38 | t(13) = -4.32 p=0.0008 d=-2.4 | t(10) = 1.86 p=0.092 d=1.3 | t(11) = -2.15 p=0.0543 d=-1.35 | t(11) = 2.6 p=0.0248 d=1.64 | t(7) = 3.27 p=0.0136 d=2.53 | t(13) = -1.5 p=0.158 d=-0.83 | t(8) = 3.04 p=0.0161 d=2.15 |
| Fatigue-Vigor | t(14) = -5.78 p=0.0 d=-3.09 | t(11) = -10.16 p=0.0 d=-6.26 | t(13) = -7.91 p=0.0 d=-4.4 | t(10) = -6.8 p=0.0 d=-4.55 | t(11) = -1.79 p=0.1002 d=-1.12 | t(11) = 2.87 p=0.0153 d=1.81 | t(7) = 2.0 p=0.0858 d=1.53 | t(13) = -0.59 p=0.5641 d=-0.33 | t(8) = 2.99 p=0.0175 d=2.11 |
| Fatigue-Tension | t(14) = -0.28 p=0.7811 d=-0.15 | t(11) = -5.14 p=0.0003 d=-3.15 | t(13) = 4.22 p=0.001 d=2.36 | t(10) = 4.78 p=0.0007 d=3.37 | t(11) = -1.04 p=0.3192 d=-0.65 | t(11) = 2.66 p=0.022 d=1.68 | t(7) = 4.58 p=0.0025 d=3.55 | t(13) = 3.75 p=0.0024 d=2.09 | t(8) = 2.54 p=0.0345 d=1.8 |
| Fatigue-Confusion | t(14) = -5.98 p=0.0 d=-3.2 | t(11) = -5.84 p=0.0001 d=-3.58 | t(13) = -8.63 p=0.0 d=-4.8 | t(10) = -6.98 p=0.0 d=-4.58 | t(11) = -4.69 p=0.0007 d=-2.93 | t(11) = 5.58 p=0.0002 d=3.54 | t(7) = -2.88 p=0.0238 d=-2.17 | t(13) = -3.65 p=0.0029 d=-2.03 | t(8) = 6.07 p=0.0003 d=4.29 |
| Vigor-Vigor | t(14) = -1.56 p=0.1405 d=-0.84 | t(11) = -1.57 p=0.1446 d=-0.96 | t(13) = -5.81 p=0.0001 d=-3.23 | t(10) = -2.22 p=0.0511 d=-1.53 | t(11) = -2.7 p=0.0208 d=-1.66 | t(11) = -4.0 p=0.0021 d=-2.48 | t(7) = -1.19 p=0.2716 d=-0.92 | t(13) = -4.28 p=0.0009 d=-2.38 | t(8) = -1.67 p=0.1329 d=-1.18 |
| Vigor-Tension | t(14) = 3.29 p=0.0054 d=1.76 | t(11) = 0.4 p=0.6947 d=0.25 | t(13) = 5.8 p=0.0001 d=3.24 | t(10) = 9.34 p=0.0 d=6.53 | t(11) = 9.22 p=0.0 d=5.73 | t(11) = 4.63 p=0.0007 d=2.86 | t(7) = 7.48 p=0.0001 d=5.79 | t(13) = 3.48 p=0.0001 d=1.94 | t(8) = -4.54 p=0.0019 d=-3.21 |
| Vigor-Confusion | t(14) = -3.48 p=0.0036 d=-1.86 | t(11) = -1.04 p=0.3221 d=-0.65 | t(13) = -3.28 p=0.006 d=-1.83 | t(10) = -6.54 p=0.0001 d=-4.46 | t(11) = -3.61 p=0.0041 d=-2.28 | t(11) = 0.13 p=0.8986 d=0.08 | t(7) = -4.4 p=0.0031 d=-3.36 | t(13) = -0.57 p=0.5757 d=-0.32 | t(8) = 2.09 p=0.0699 d=1.48 |
| Tension-Tension | t(14) = 3.73 p=0.0023 d=1.99 | t(11) = -6.41 p=0.0 d=-4.0 | t(13) = 4.26 p=0.0009 d=2.38 | t(10) = 5.39 p=0.0003 d=3.8 | t(11) = 3.44 p=0.0055 d=2.18 | t(11) = -2.47 p=0.0312 d=-1.54 | t(7) = 5.0 p=0.0016 d=3.87 | t(13) = 2.19 p=0.047 d=1.22 | t(8) = -3.48 p=0.0083 d=-2.46 |
| Tension-Confusion | t(14) = 3.27 p=0.0056 d=1.75 | t(11) = -2.15 p=0.0544 d=-1.32 | t(13) = 4.33 p=0.0008 d=2.42 | t(10) = 3.77 p=0.0037 d=2.57 | t(11) = 6.5 p=0.0 d=4.04 | t(11) = 3.76 p=0.0032 d=2.32 | t(7) = 5.21 p=0.0012 d=4.03 | t(13) = 1.74 p=0.106 d=0.97 | t(8) = -3.06 p=0.0157 d=-2.16 |
| Confusion-Confusion | t(14) = 1.26 p=0.23 d=0.67 | t(11) = -1.62 p=0.133 d=-0.99 | t(13) = -1.11 p=0.2881 d=-0.62 | t(10) = -3.8 p=0.0035 d=-2.54 | t(11) = 2.61 p=0.0241 d=1.63 | t(11) = 3.01 p=0.0118 d=1.87 | t(7) = -3.28 p=0.0135 d=-2.52 | t(13) = -2.36 p=0.0345 d=-1.31 | t(8) = 0.1 p=0.9241 d=0.07 |

· Abbreviation of datasets: EQ2011(non-EQ) - Earthquake2011 (not including earthquake keywords), EQ2011- Earthquake2011 (filtered by earthquake keywords), EQ2018(non-EQ) - Earthquake2018 (not including earthquake keywords), EQ2018- Earthquake2018 (filtered by earthquake keywords), BS: Bluesky

# 7 Alternative approach to calculate the emotion link's strength by summing up the strength of significant concept links.

In the manuscript, we build emotion networks by counting the number of significant concept links between a pair of emotions as the strength of emotion links. Here, we explore an alternative approach-summing up the strength of concept links. The strength of concept links is measured by the ratio between the actual occurrence of a concept link in real data and the random shuffled case (Equation 2).

In Fig. S9a, we compare the strength of emotion links by counting the number of significant concept links (x-axis) and summing the strength of significant concept links (y-axis). When summing the strength of significant concept links, we excluded outliers—concept link strengths identified as outliers using a boxplot. In Fig. S9a, a clear linear relationship can be observed between the two methods, thus supporting our conclusion in the main text.

Furthermore, in Fig. S9b and Fig. S9c, we present the Spearman correlations between emotion networks within each dataset across time (Fig. S9b) and across different datasets (Fig. S9c), respectively. The sub-figures on the left show comparisons of emotion networks based on the count of significant concept links (the method used in the main manuscript), while the sub-figures on the right show comparisons based on the sum of the strengths of significant concept links. Both the within-dataset and across-dataset comparisons yield consistent results that support our main conclusions: emotion networks exhibit high stability over time, with only minor variations observed before and after the start of vaccination (Fig. S9b); datasets collected during crises show strong similarity to those from normal periods, and the earthquake-related datasets show greater similarity to the pre-vaccination datasets than to the non-earthquake or vaccination-related datasets.

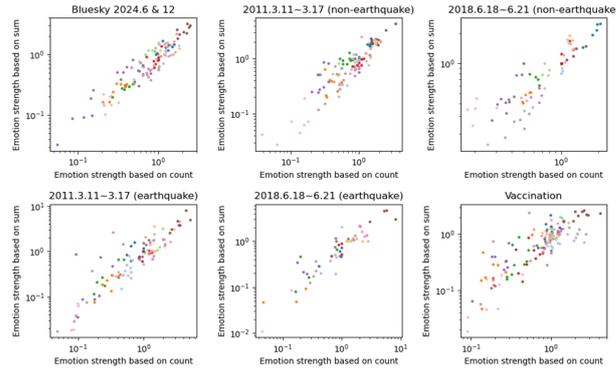

(a) Emotion link strength based on count of significant links vs. sum of strength of significant links

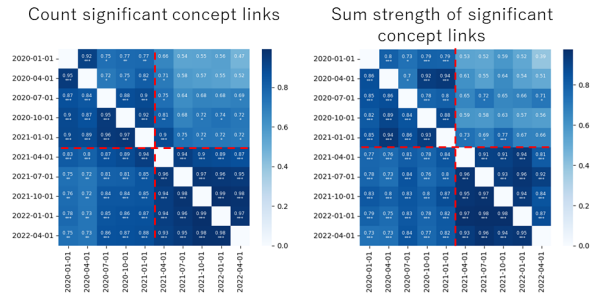

(b) Spearman correlation of emotion links within dataset (vaccination)

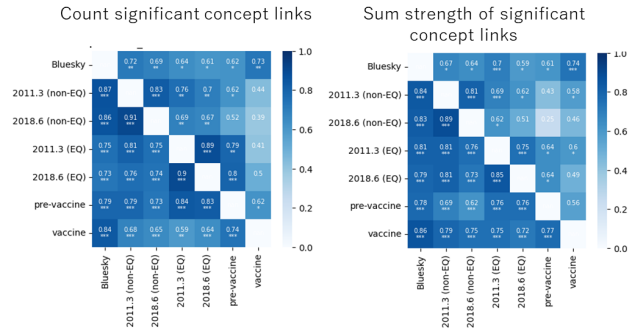

(c) Spearman correlation of emotion links across datasets

**Supplementary Fig. S9**: Measure the strength of emotion links by summing up the strength of concept links between and within emotions. (a) Scatterplot comparing the strength of emotion links based on counting of significant concept links vs. summing up the strength of significant concept links. Each dot in the figure represents the strength of an emotion link within a time window. (b) Spearman correlation of emotion networks within dataset across time, based on count (left) or sum (right) of strength of significant links. (c) Spearman correlation of emotion networks across datasets, based on count (left) or sum (right) of strength of significant links.

# 8 Sensitivity analysis

We evaluate the sensitivity of our results through two analyses. First, we examine how the findings vary with different thresholds for selecting statistically significant emotion co-occurrence links. Second, we assess the robustness of the results by randomly sampling the same number of words from each emotion dimension in the POMS. These analyses allow us to test the reliability of our approach under varying methodological assumptions.

**(1) Altering thresholds for filtering significant concept links**

In SI Fig. S10, we replicate the analysis presented in **Fig. 3**, which compares emotion networks across datasets, using alternative threshold settings. As shown in SI Fig. S10a, the overall emotion networks—including both intra- and inter-emotion links—remain significantly similar across datasets, which is aligned to **Fig. 3a**. However, the inter-emotion networks exhibit greater variability from **Fig. 3b**, indicating that the similarity of inter-emotion networks is more sensitive to threshold changes. In the following, we investigate in details how does the similarity of inter-emotion network change based on different threshold, by comparing to **Fig. 3b**.

Based on **Fig. 3b**, we drew four key conclusions related to the Spearman's rank correlation between datasets:

- Emotion networks during crisis show significant similarity to the normal time
- The similarities within earthquake-related datasets are significantly
- The similarities within vaccination-related datasets are significantly
- The similarities between earthquake and vaccination are not significant

In SI Fig. S11 we evaluate if these four key conclusions remain valid under different threshold settings, by examining the average Spearman's $p$-value between datasets. We observe that when the weight threshold $W \leq 30\%$ (in the example shown in SI Fig. S14a, top 30% of weight is equivalent to more than 5 occurrences) and the strength threshold $S \geq 1$, the conclusions generally hold. However, if we further relax these thresholds to include more randomly or infrequently used concept links, these conclusions may no longer remain robust.

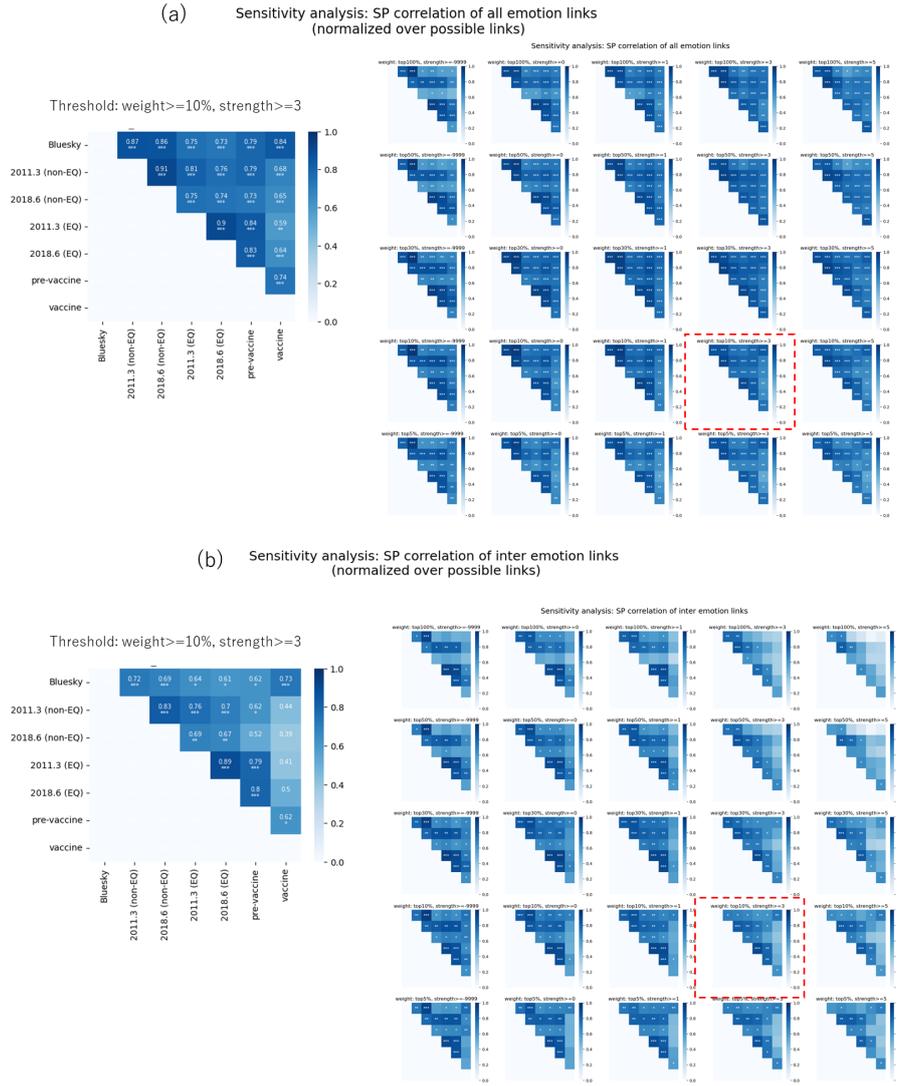

· Asterisks: *: $p$-value $\leq 0.05$, **: $p$-value $\leq 0.01$, ***: $p$-value $\leq 0.001$.

**Supplementary Fig. S10**: Heatmap showing detailed Spearman's rank correlation between datasets for different thresholds. (a) Spearman correlation of emotion links including inter and intra-emotion links. (b) Spearman correlation of emotion links including inter-emotion links. In all sub-figures, the y-axis and x-axis follow the heatmap example shown on the left. The annotated asterisks represent the significance of Spearman correlation, adjusted based on False Discovery Control.

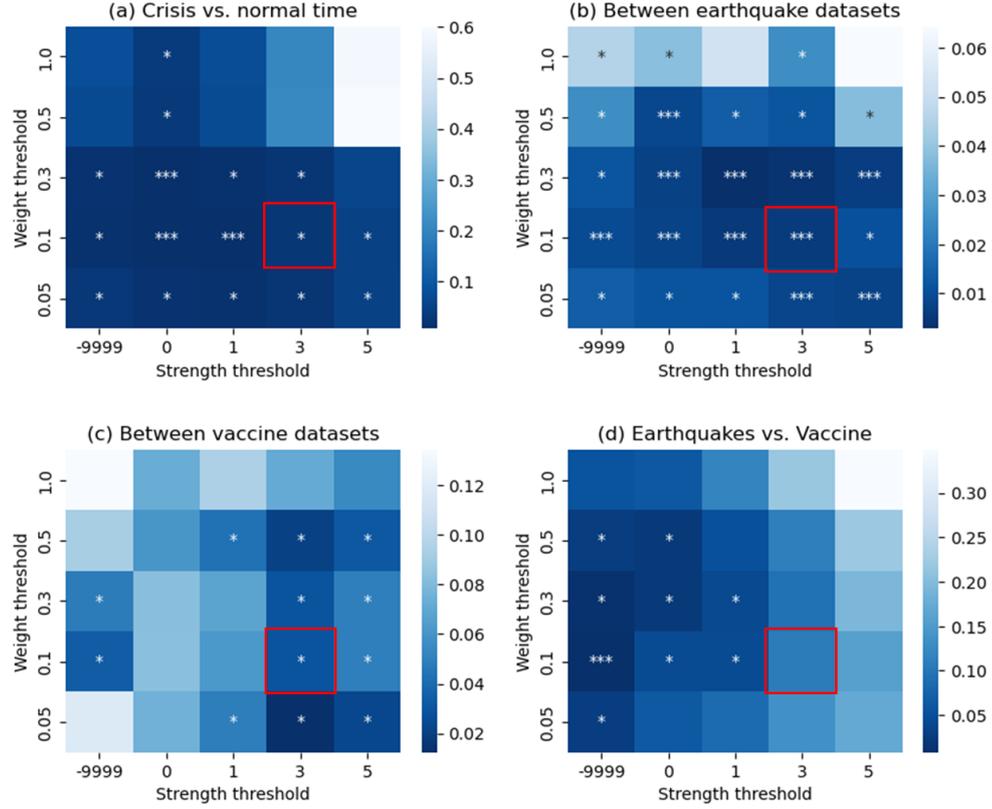

· Asterisks: *: $p$-value $\leq 0.05$, **: $p$-value $\leq 0.01$, ***: $p$-value $\leq 0.001$.

**Supplementary Fig. S11**: Summary of the average $p$-value of inter-emotion network similarity for different thresholds. (a) Spearman's $p$-value between crisis datasets and normal-time dataset. (b) Spearman's $p$-value between earthquake datasets. (c) Spearman's $p$-value between vaccine datasets. (d) Spearman's $p$-value between vaccination versus earthquake datasets. The annotated asterisks represent the level of significance based on the average $p$-value calculated from Fig. A7. The bins that are circled in red represent the thresholds adopted in the Result section.

**(2) Sampling the same number of concepts (words) from each emotion dimension**

Next, we evaluate the impact of unbalanced word (concept) counts across emotion dimensions on our results. As described in the Methods section, the number of words per dimension in the POMS dictionary ranges from 70 (Fatigue) to 197 (Confusion). While we accounted for this imbalance in the main analysis by normalizing the number of significant concept links over the total possible links between two emotion

20

dimensions, we further conduct a sensitivity analysis to assess the robustness of our findings.

Specifically, we draw a hundred samples from the POMS dictionary, each time randomly drawing the same number of words ($n = 70$) per emotion dimension. For each sample, based on the previously calculated significance and weight, we filter the concept links that are above the thresholds and contain the selected words. We then replicate the analyses presented in **Fig. 2** and **Fig. 3**, comparing the Spearman's correlation coefficients and corresponding $p$-values from the original data with those obtained from the resampled networks.

In SI Fig. S12, we compare the similarity of inter-emotion networks within and across datasets between the actual data and the randomly sampled data. The black error bars represent the median and the 95% confidence interval of the Spearman correlation coefficients and corresponding $p$-values computed from randomly sampled networks. We can observe that the randomly sampled dataset does not change significantly from the actual data (actual data falls within random data's 95% confidence interval). Even though some random samples show an upper bound of $p$-value exceeding the 0.05 significance threshold (i.e. Bluesky within dataset similarity), in general, our results are robust.

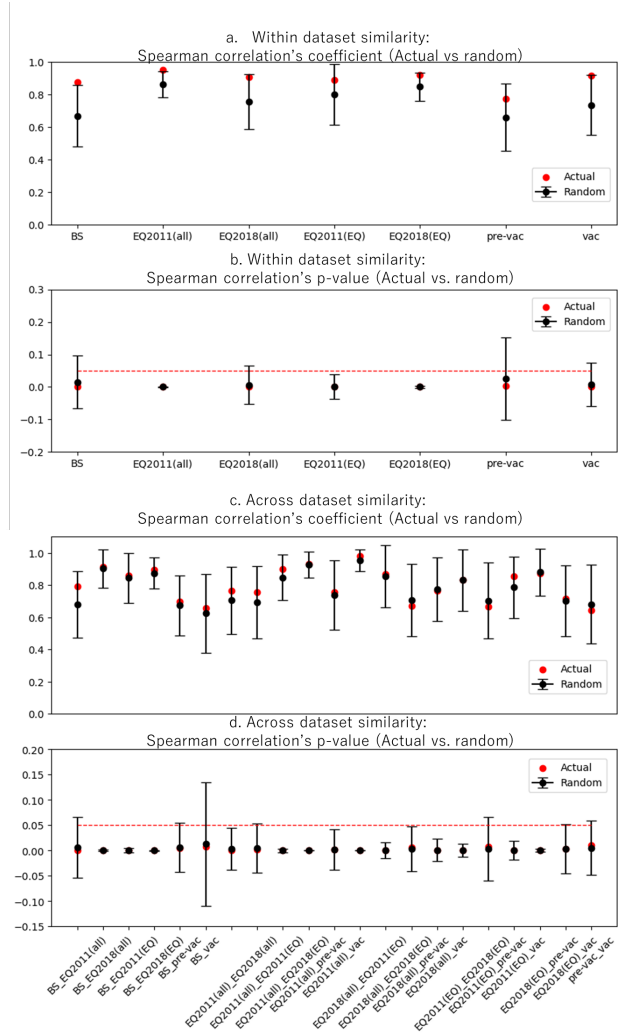

**Supplementary Fig. S12**: Comparing the similarity of inter-emotion networks between the actual data and the randomly sampled data. (a) Spearman correlation coefficient of the temporal emotion networks within each dataset. (b) Spearman correlation's $p$-value of the within dataset comparison. (c) Spearman correlation coefficient of emotion networks across dataset. (d) Spearman correlation's $p$-value of the across dataset comparison. In all sub-figures, the red dots depict the actual data. The error bar in black represents the randomly sampled data, showing median with 95% confidence interval.

# 9 Calculate concept link's significance

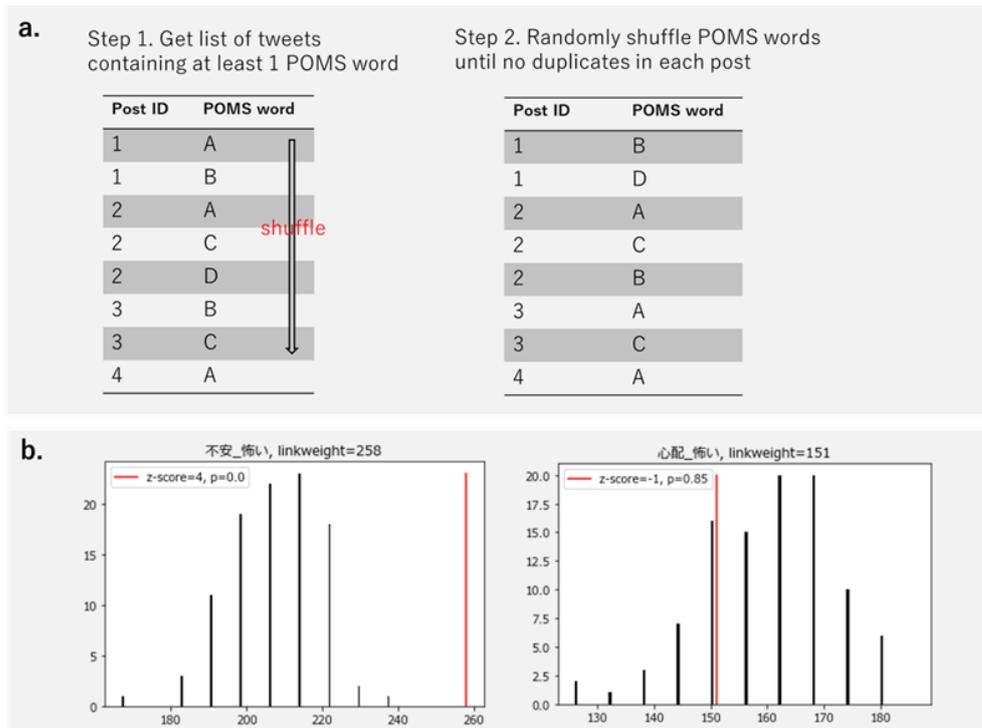

**Supplementary Fig. S13**: Calculate concept link's significance. (a) Process of shuffling to create random concept networks. (b) Example of word pairs with high (left) or low (right) significance. The black bars show the histogram chart of link weight based on random networks, while the red line shows the actual link weight represented by the number of tweets in which the pair of words co-occurred.

# 10 Filter significant links based on weight and strength thresholds.

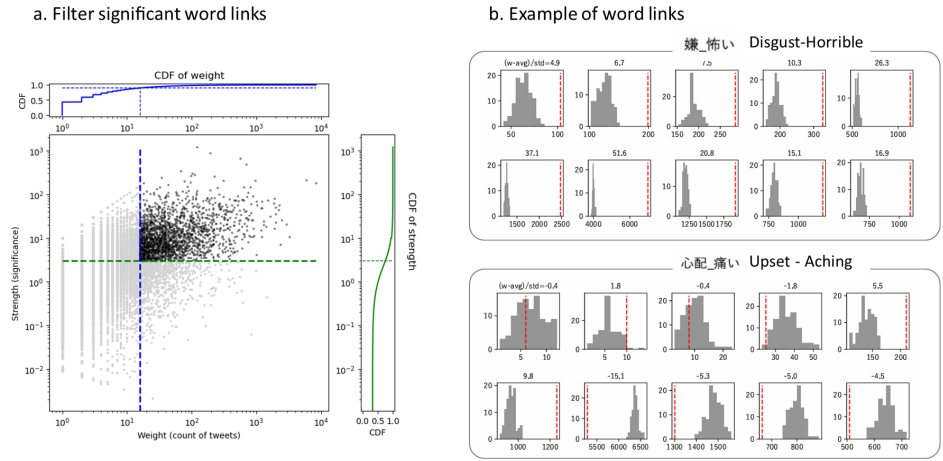

**Supplementary Fig. S14**: Filter significant links based on weight and strength thresholds. (a) Scatterplot of weight and strength of concept links. The CDF of weight horizontally aligned to the x-axis shows that based on this dataset when we adopt a weight threshold W $\geq 10\%$, it is equivalent to link weight $\geq 15$. The CDF of strength vertically aligned to the y-axis depicts that we adopt the strength threshold $S \geq 3$. Each node in the scatterplot represents a concept link, and those in black are the ones above both thresholds and therefore are regarded as significant links. (b) Example of concept links with high occurrence frequency but different significance stability. There are 10 subplots for each example, representing 10 quarters of vaccination datasets. In each subplot, the grey histogram depicts the distribution of random weight from a hundred shuffling. The red dashed line represents the actual link weight. The first concept link (disgust-horrible) displays a consistent high significance while the second concept link (horrible-tough) displays fluctuating significance



\end{document}